\documentclass[trackchanges]{aastex7}

\usepackage{subcaption}
\usepackage{graphicx}
\usepackage{amsmath}

\begin{document}

\title{Validating DIRECD: Statistical Evaluation of Coronal Mass Ejections Direction Estimates from Coronal Dimmings}

\author[orcid=0000-0003-3883-8960]{Shantanu Jain}
\affiliation{Center for Digital Engineering, Bolshoy Boulevard 30, bld. 1, Moscow 121205, Russia}
\email[show]{s.jain@rcdei.com}  

\author[orcid=0000-0002-9189-1579]{Tatiana Podladchikova} 
\affiliation{Center for Digital Engineering, Bolshoy Boulevard 30, bld. 1, Moscow 121205, Russia}
\email{t.podladchikova@rcdei.com}

\author[orcid=0000-0001-5661-9759]{Karin Dissauer}
\affiliation{NorthWest Research Associates, 3380 Mitchell Lane, Boulder, CO 80301, USA}
\email{dissauer@nwra.com}

\author[orcid=0000-0003-2073-002X]{Astrid M. Veronig}
\affiliation{University of Graz, Institute of Physics, Universitätsplatz 5, 8010 Graz, Austria}
\affiliation{University of Graz, Kanzelh\"ohe Observatory for Solar and Environmental Research, Kanzelh\"ohe 19, 9521 Treffen, Austria}
\email{astrid.veronig@uni-graz.at}

\author[0000-0003-2735-5832]{Amaia Razquin}
\affiliation{University of Graz, Institute of Physics, Universitätsplatz 5, 8010 Graz, Austria}
\email{amaia.razquin-lizarraga@uni-graz.at}


\begin{abstract}

Coronal mass ejections (CMEs) are among the most energetic phenomena in our solar system, with significant implications for space weather. Understanding their early dynamics remains challenging due to observational limitations in the low corona. We present a statistical evaluation of the DIRECD (Dimming InfeRred Estimation of CME Direction) method, which provides a novel approach to determining initial CME propagation directions using coronal dimmings. We analyze 33 coronal dimming events well observed by SDO/AIA and validate our DIRECD results with 3D reconstructions from the Graduated Cylindrical Shell (GCS) model. We find generally good agreement between the DIRECD-derived inclinations and the GCS model. In the meridional plane (north--south direction), the mean difference in inclinations is \(0.3^\circ \pm 7.8^\circ\). 
In the equatorial plane (east--west direction), the mean difference is \(-2.9^\circ \pm 18.9^\circ\). In 3D, the inclinations show a mean difference of \(1.2^\circ \pm 10.4^\circ\). We further visually compare our method by projecting the DIRECD cones onto LASCO/C2 observations, and verify the model's ability to capture both the primary CME structure and associated secondary dimming regions. This work establishes DIRECD as a powerful, observationally grounded technique for determining the initial CME direction, offering new insights that complement existing reconstruction methods. The technique's unique capability to determine early CME direction in the low corona using coronal dimmings observed in EUV images makes it particularly valuable for improving space weather forecasting models.
\end{abstract}



\section{Introduction}

Coronal Mass Ejections (CMEs) are massive clouds of magnetized plasma expelled from the Sun to interplanetary space with speeds of some hundreds to thousands of km/s \citep{michalek2009expansion, gopalswamy2009soho, tsurutani2014extreme, Cheng2017, Veronig2018_Genesis}. They can have severe effects on our technology, including disrupting radio transmissions, inducing harmful currents in power grid systems, and lowering the orbits of satellites orbiting the Earth \citep{sandford1999impact,doherty2004space,baker2013major, parker2024satellite}. Therefore, investigating and identifying the initial development of CMEs is crucial. Due to projection effects, coronagraphs do not provide clear observations of their evolution \citep{burkepile2004role,schwenn2005association}, and their origin remains unknown due to the occultation disk blocking the lower corona. As a result, numerous studies have examined whether any phenomenon associated with CMEs could offer a deeper understanding of their early stages. One such phenomenon are coronal dimmings that form due to an expansion and evacuation of plasma in the low corona. They are observed as localized reductions in extreme ultraviolet (EUV) and soft X-ray emissions in the low corona, and are closely related to CMEs \citep{hudson1996long, sterling1997yohkoh, thompson1998soho, 2025LRSP...22....2V}.

Several studies have demonstrated how coronal dimmings relate to the mass of CME(e.g. \citep{harrison2000spectroscopic, harrison2003coronal, zhukov2004nature, lopez2017mass}, 
the morphology and early evolution of the CME \citep{attrill2006using, qiu2017gradual}, and their timing \citep{miklenic2011coronal}. Furthermore, multiple studies have investigated the statistical correlation between CMEs and coronal dimmings \citep{bewsher2008relationship, reinard2009relationship, mason2016relationship, krista2017statistical, aschwanden2017global,razquin2025coronal}. These studies have identified a strong relationship between CME mass and dimming characteristics, including dimming area and brightness. Additionally, several studies have examined connections between the CME's maximum speed and other key parameters such as rate of area expansion, brightness variation, magnetic flux evolution, and the average intensity of dimming regions \citep{Dissauer2018a,Dissauer2018b, Dissauer2019,chikunova2020coronal}.  Furthermore, dimmings have also been suggested as a potential proxy for stellar CMEs \citep{Jin2020coronal, Veronig2021indications}.

Recent studies suggest that the expansion and morphology of dimmings provide valuable insights into the CME expansion and recovery processes \citep{ronca2024recovery}, as well as investigating CME propagation directions and deflections using coronagraph data \citep{mostl2015strong,  jain2024coronal,jain2024estimating, podladchikova2024three}. 

The advent of the STEREO mission enabled multi-viewpoint heliospheric observations, facilitating the development of various 3D reconstruction techniques for CME geometry and trajectory \citep{mierla20103,isavnin2013three,liewer2015observations,kay2017deflection}. Recent work by \citet{Chikunova2023} established a correlation between coronal dimming morphology and Graduated Cylindrical Shell (GCS) fitted CME structures, suggesting that dimming observations could serve as a diagnostic tool for determining the CME direction. This concept has been further developed into the DIRECD method, which estimates the initial CME propagation direction using coronal dimmings \citep{jain2024coronal, jain2024estimating, podladchikova2024three}. While DIRECD focuses on early-stage CME direction, techniques GCS model \citep{thernisien2006modeling, thernisien2011implementation} provide 3D reconstructions of CMEs once they are observed in coronagraphs. We present a statistical analysis of the DIRECD method to estimate early CME direction from the spatio-temporal evolution of coronal dimmings in the low corona. The DIRECD method is applied to a statistical set of 33 events, where the dimming is observed against the solar disk by EUV images.

\section{Data Set} \label{sec:dataset}


This study utilizes full-disk EUV imagery from the Atmospheric Imaging Assembly \citep[AIA;][]{lemen2012atmospheric} onboard the Solar Dynamics Observatory \citep[SDO;][]{pesnell2012solar}. As demonstrated by \cite{Dissauer2018a} and \cite{kraaikamp2015solar}, coronal dimmings are most effectively observed in wavelength bands sensitive to quiet Sun coronal temperatures, particularly at 195, 171, and 211~\AA. For the present analysis, the AIA 211~\AA~filter is employed, with a temporal resolution (cadence) of one minute per observation. For each event under investigation, a sequence of images is analyzed, commencing 30 minutes prior to the onset of the associated solar flare as determined from the HINODE flare catalog \citep{watanabe2012hinode}  and extending over a total duration of 4 hours. All SDO images within the dataset were checked for exposure time ($>1$ second), corrected for differential rotation, and resampled to a uniform resolution of 1024 × 1024 pixels using the SunPy library in Python \citep{mumford2015sunpy}. The detection of coronal dimming regions was confined to a 1000 × 1000 arc second subfield centered on the eruption site.

We combine two complementary datasets: (1) the dimming catalog of \cite{Dissauer2018b} (62 events from 2010--2012) and (2) the CME catalog of \cite{kay2017using} (45 events from 2010--2014). Both catalogs overlap during the STEREO era, leveraging multi-spacecraft observations. After merging these catalogs and removing duplicate events, we visually inspected each event and identified 33 events suitable for DIRECD analysis in terms of the following criteria:


\begin{itemize}
    \item Clear and Persistent Dimming Signatures:  Events must exhibit unambiguous coronal dimming in SDO/AIA 211 Å base difference images. Faint, patchy, or transient dimmings were excluded.
    \item Well-defined CME Association: Each dimming must be associated with an observed CME, with a clear onset time and kinematic properties recorded in the merged catalog.
    \item Flare Association: All selected events must be associated with flare, as verified through the GOES flare catalog\footnote{http://www.ngdc.noaa.gov/stp/satellite/goes/index.html} and/or HINODE flare catalog \citep{watanabe2012hinode}.
    \item Flare source location: The flare source region must lie within $\pm~ 60~^\circ$  of disk center to minimize projection errors in the dimming identification.
    
\end{itemize}

We excluded two previously analyzed events (1~October~2011 and 6~September~2011) that were already examined in detail by \cite{jain2024coronal}, while augmenting our sample with two recent scientifically significant events: the 2019 eruption studied by \cite{dumbovic20212019} and the 2021 event discussed in \cite{Chikunova2023}.

\section{Methods and Analysis} \label{sec:method}

\subsection{Dimming Detection}

The detection of coronal dimmings is performed following the methodology outlined in \cite{Dissauer2018a, Chikunova2023}. The approach employs a region-growing and thresholding algorithm applied to Logarithmic Base Ratio (LBR) images, with the detection threshold set between $-0.11$ DN and $-0.15$ DN , adjusted based on dimming intensity. To minimize noise and spurious pixel detections, morphological operations (using a $3\times3$ pixel kernel) are applied to smooth the identified regions and eliminate small-scale artifacts, supplemented by median filtering.

From this process, we derive both instantaneous dimming maps (all dimming pixels detected for one time step) and cumulative dimming maps (dimming pixels detected over the entire observational period). Additionally, the end of the impulsive phase of the dimming is determined following \cite{Dissauer2018b}, who defined it as the time at which the growth rate of the cumulative dimming area (dA/dt) declines to 15\% of its peak value. Table \ref{table:events} shows the start/peak/end time of the dimming impulsive phase and the flare source.

\begin{table}[h]
\caption{List of selected dimming events.}

\begin{tabular}{|l|l|l|l|l|c|c|}
\hline
\textbf {\#} & \textbf{Event Date} & \textbf{Start Time (UT)} & \textbf{Peak Time (UT)} & \textbf{End Time (UT)} & \textbf{Lat ($^\circ$)} & \textbf{Lon ($^\circ$)} \\  \hline
1& 16-07-2010 & 14:46:00 & 15:17:00 & 15:43:00 & $-21$ & $20$  \\ \hline
2&01-08-2010 & 07:03:00 & 07:45:00 & 08:17:00 & 20  & $-35$ \\ \hline
3&07-08-2010 & 17:35:00 & 18:15:00 & 18:39:00 & 14  & $-37$ \\ \hline
4&13-02-2011  & 17:00:36 & 17:58:36 & 18:33:36 & $-20$ & $-5$  \\ \hline
5&15-02-2011  & 01:20:00 & 02:21:00 & 02:53:00 & $-20$ & 12  \\ \hline
6&07-03-2011 & 13:16:00 & 14:09:00 & 14:58:00 & 12  & $-21$ \\ \hline
7&21-06-2011 & 00:52:36 & 02:29:36 & 03:25:36 & 14  & 10  \\ \hline
8&02-08-2011 & 05:58:00 & 06:38:00 & 06:56:00 & 15  & 14  \\ \hline
9&03-08-2011 & 12:54:00 & 14:13:00 & 14:58:00 & 17  & 30  \\ \hline
10&06-09-2011 & 01:35:00 & 02:20:00 & 02:51:00 & 14  & 7   \\ \hline
11&27-09-2011 & 20:44:00 & 21:10:34 & 21:30:29 & 10  & $-9$  \\ \hline
12&26-11-2011 & 06:09:00 & 06:53:00 & 07:39:00 & 11  & 47  \\ \hline
13&25-12-2011 & 18:11:00 & 18:22:00 & 19:23:00 & $-22$ & 26  \\ \hline
14&26-12-2011 & 11:23:00 & 11:23:57 & 11:53:24 & 20  & 1   \\ \hline
15&26-12-2011 & 01:33:07 & 02:27:07 & 02:59:07 & $-18$ & 30  \\ \hline
16&19-01-2012 & 13:43:00 & 15:02:00 & 15:54:00 & 32  & $-27$ \\ \hline
17&07-03-2012 & 23:34:12 & 00:28:12 & 00:59:12 & 18  & $-31$ \\ \hline
18&09-03-2012 & 03:00:48 & 03:46:48 & 04:07:48 & 17  & $-2$  \\ \hline
19&10-03-2012 & 16:46:48 & 17:42:48 & 18:10:48 & 16  & 24  \\ \hline
20&05-04-2012 & 20:49:00 & 20:53:00 & 21:18:00 & 24  & 32  \\ \hline
21&03-06-2012 & 17:27:00 & 18:03:00 & 18:28:00 & 17  & $-38$ \\ \hline
22&06-06-2012 & 19:26:48 & 20:22:48 & 20:47:48 & $-18$ & 5   \\ \hline
23&14-06-2012 & 12:23:36 & 14:01:36 & 14:39:36 & $-17$ & $-5$  \\ \hline
24&04-07-2012 & 16:09:23 & 16:41:23 & 17:26:23 & 12  & 35  \\ \hline
25&25-09-2012 & 04:05:35 & 04:27:35 & 04:48:35 & 9   & $-20$ \\ \hline
26&11-04-2013 & 06:55:11 & 07:21:11 & 08:24:11 & 9   & $-12$ \\ \hline
27&17-05-2013 & 08:15:23 & 09:00:23 & 10:06:23 & 12  & $-42$ \\ \hline
28&06-10-2013 & 13:37:11 & 14:02:11 & 14:41:11 & $-15$ & 12  \\ \hline
29&10-11-2013 & 05:07:59 & 05:17:59 & 05:42:59 & $-14$ & 14  \\ \hline
30&18-04-2014 & 06:21:59 & 07:08:59 & 07:54:59 & $-23$ & 39  \\ \hline
31&20-12-2014 & 23:43:59 & 00:34:59 & 01:34:59 & $-21$ & 24  \\ \hline
32&08-03-2019 & 02:38:57 & 03:12:57 & 03:33:57 & $9$   & 3   \\ \hline
33&28-10-2021 & 14:59:09 & 16:00:09 & 16:22:09 & $-28$ & 1   \\ \hline
\end{tabular}
\tablecomments{We list here the date, start/peak/end time of the dimming impulsive phase based on our detection, and the heliographic location of the associated flare.}
\label{table:events}
\end{table}

\begin{figure}
\centering 
\subfloat[]{%
 \includegraphics[width=0.32\columnwidth]{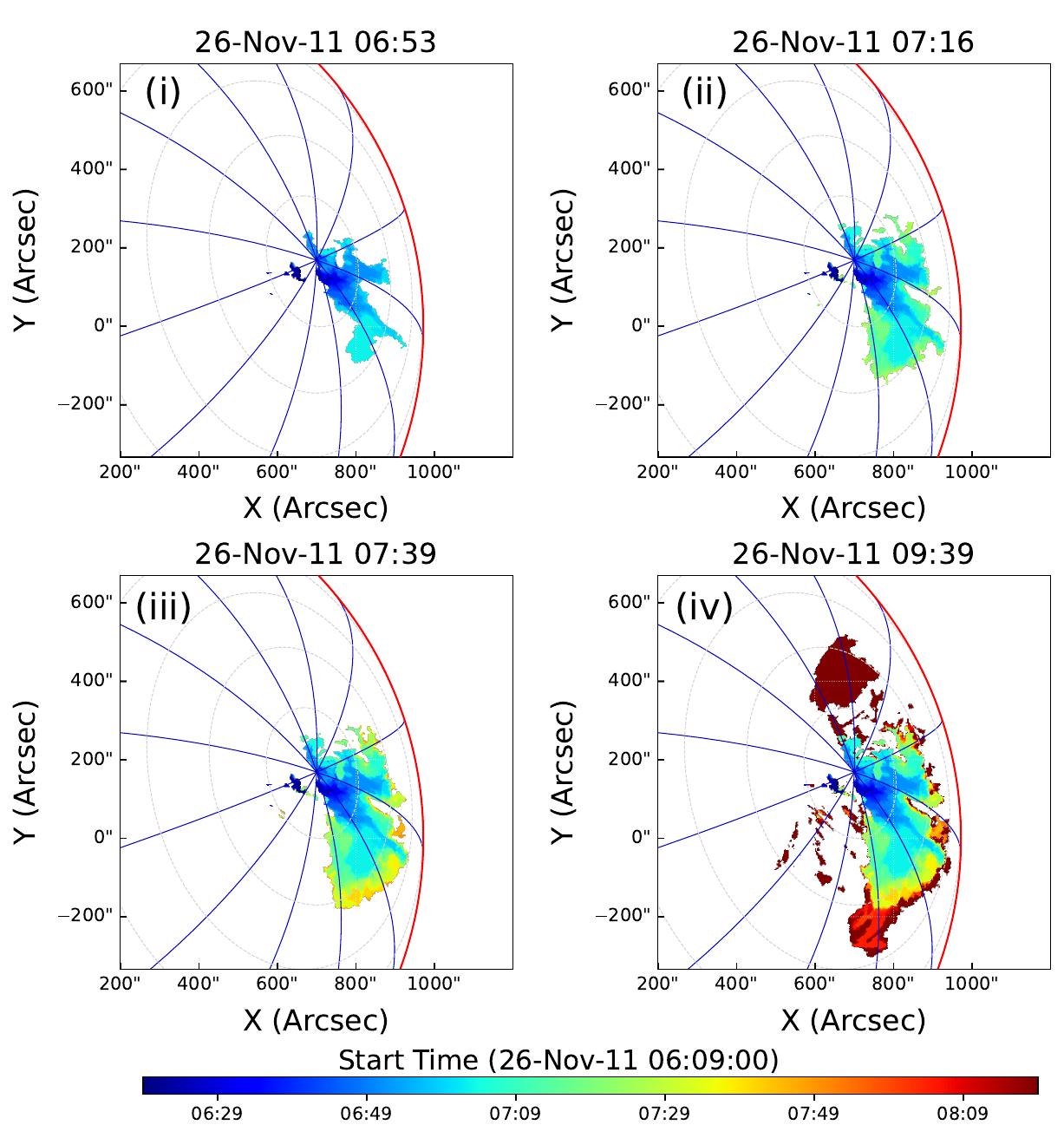}%
  \label{fig:dimming_nov}%

}
\subfloat[]{%
  \includegraphics[width=0.65\columnwidth]{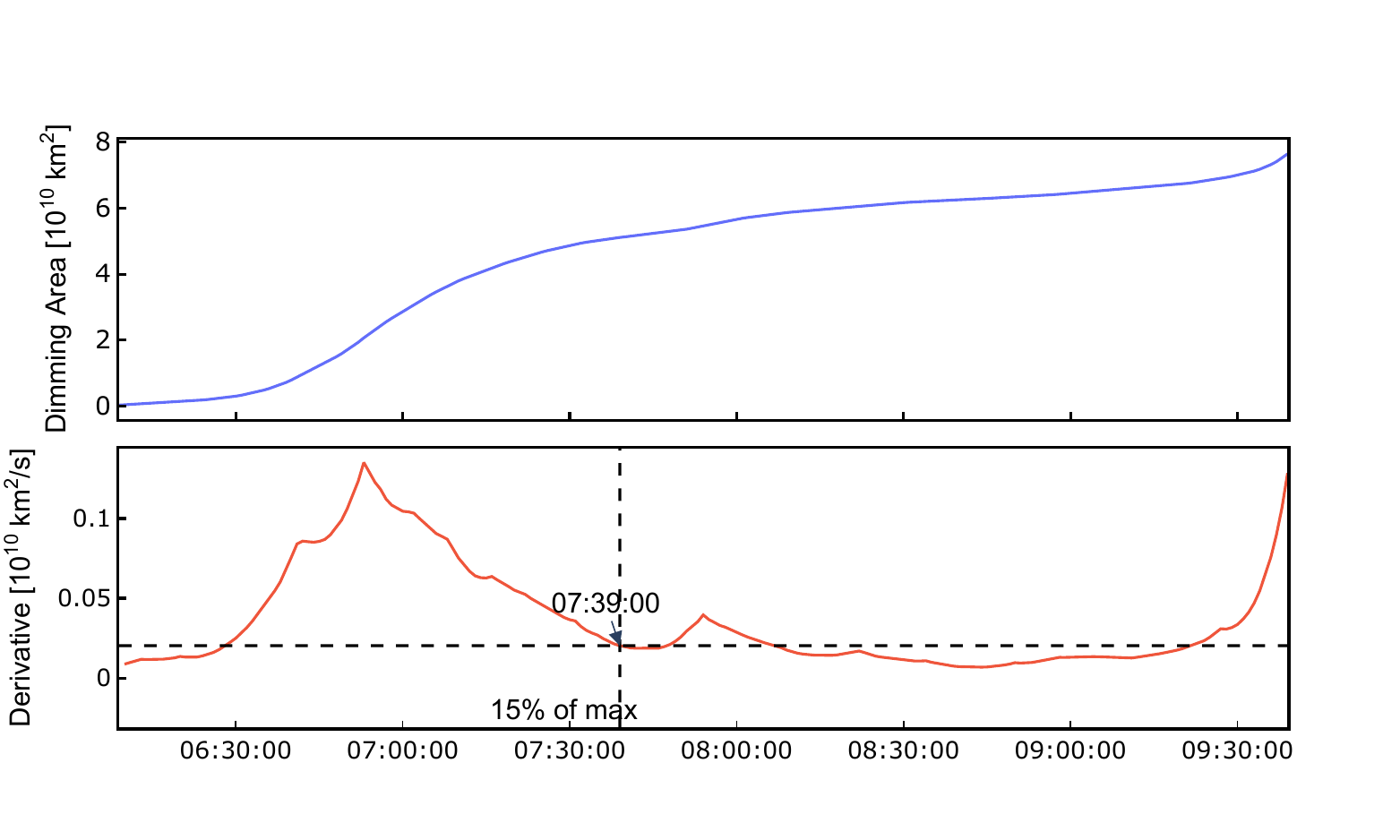}%
  \label{fig:dimming_der_nov}%

}
\caption{Panel (a): Dimming evolution for the November 26, 2011 event at (i) the maximum of the impulsive phase (reached 44 min after its start); (ii) 67 min after the start of the event; (iii) at the end of the impulsive phase (90 min after the start); and (iv) 3.5 h after the start. The color bar shows when each dimming pixel was detected for the first time. The blue lines indicate 12 angular sectors to determine the dominant dimming direction and the red curve indicates the solar limb. Panel (b): Dimming area expansion A (top panel) and it's derivative $dA/dt$ (bottom panel) for the November 26, 2011 event. The end of the impulsive phase is marked by a vertical dotted line at 07:39 UT when $dA/dt$ falls below 15\% of its maximum value.}
\end{figure}

Figure~\ref{fig:dimming_der_nov} shows the dimming area expansion and end of the impulsive phase for an example event \#12 on November 26, 2011 and Figure~\ref{fig:dimming_nov} shows the cumulative dimming maps at different time steps across the three hour period. Each individual pixel of the presented cumulative dimming pixel mask is colored according to when it was first detected from the start time of the flare/CME from HINODE. Blue dimming regions were detected at the beginning of the event, whereas the red regions were detected later. In this figure, panel c shows the dimming map at the end of the impulsive phase at 07:39~UT (90 min after the start).

\subsection{Application of DIRECD}

To apply the DIRECD method on our selected events, we follow the methodology described in \citep{jain2024coronal,jain2024estimating}. Firstly, we perform a sector-based analysis, dividing the solar sphere centered at the flare source into 12 sectors of $30^\circ$ each and derived the dimming area profiles A(t) in each sector. 

\begin{figure}
\centering 
\subfloat[]{%
  \includegraphics[width=0.35\columnwidth]{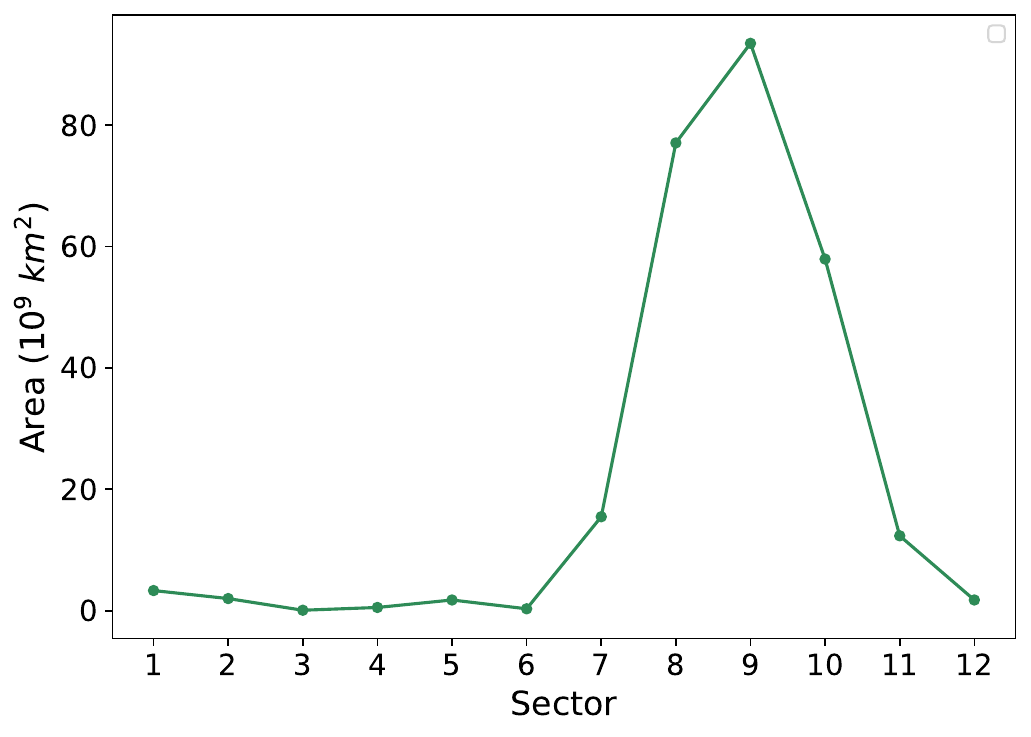}%
  \label{plot:area_sec_nov}%
}\qquad
\subfloat[]{%
  \includegraphics[width=0.5\columnwidth]{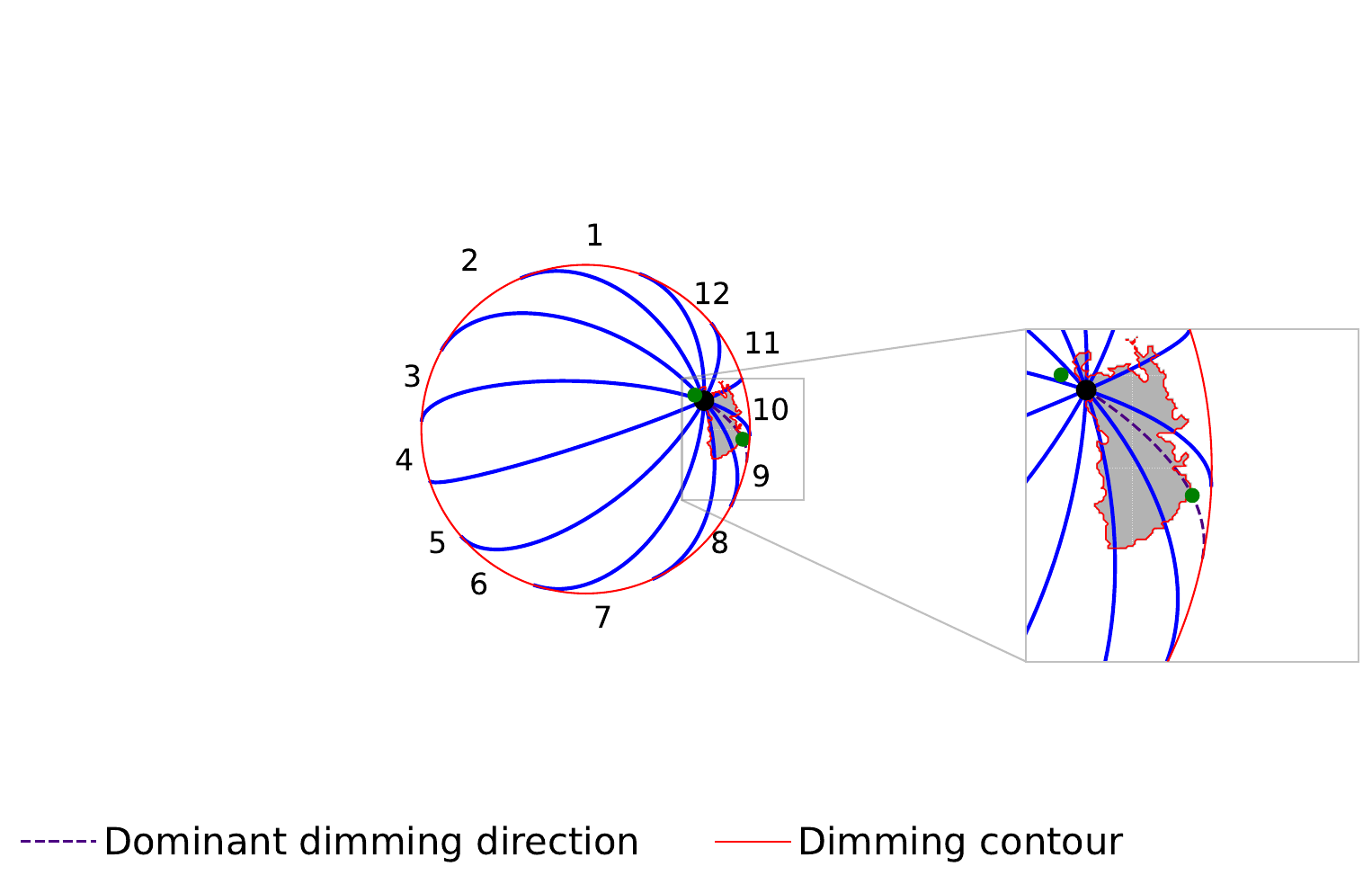}%
  \label{plot:dominant_nov}%
}
\caption{Left Panel: Area-sector plot for the November 26, 2011 event. The largest area is in sector 9. Right Panel: Dominant dimming direction in sector 9, plotted with dimming mask (in grey) and dimming edges (in green). }
\end{figure}

Figure~\ref{plot:area_sec_nov} displays the dimming area profile versus the sector number (panel a) for an example event \#12 measured at 07:39~UT on November 26, 2011, corresponding to the end of the impulsive phase. The dominant dimming direction, identified as sector 9 in Figure~\ref{plot:dominant_nov}, corresponds to the sector exhibiting the largest cumulative dimming area. 

Next, we perform dimming edge selection as outlined in \citep{jain2024coronal, jain2024estimating}. However, here we slightly modify the dimming edge selection criteria to ensure that the resulting projections of the cone consistently encompass approximately 100\% of the dimming region. To select the dimming edges, we employ a systematic approach in which one edge is identified at the largest dimming extent within the sector corresponding to the dominant dimming direction, while the second edge is determined by the maximum dimming expansion observed in the remaining sectors, excluding sectors adjacent to the dominant dimming sector. We then define two dimming edges:
\begin{itemize}
    \item Edge 1: At the largest dimming extent within the sector of dominant dimming direction (Sector 9 in figure \ref{plot:dominant_nov})
    \item Edge 2: Opposite to sector of dominant dimming direction and at a distance of largest dimming expansion from source in all sectors except sector of dominant dimming direction and its two adjacent sectors on either side i.e excluding sectors 7-11 in figure \ref{plot:dominant_nov}. The exclusion criteria was carefully chosen to cover the full extent of the dimming region while keeping the projection boundaries as close as possible to its actual morphology, providing a good trade-off between the two. This gives us the second edge along sector 2 at a distance of largest dimming extent of sector 1. 
\end{itemize}

Using the selected dimming edges, we model the CME geometry ensuring that their projections accurately match the observed dimming morphology. Each cone in the ensemble was parameterized by its unique height (H), inclination angle ($\beta$), and half angular width ($\alpha$), preserving a physically consistent link to the dimming region. A brief explanation is provided in appendix \ref{appendix1} (for full details of cone construction refer to \cite{jain2024coronal}).

Figures \ref{fig:cone_projection_1}, \ref{fig:cone_projection_2} in the Appendix~\ref{appendix2}
present an ensemble of 20 reconstructed CME cones for the example event \#12 of November 26, 2011, with geometric parameters spanning the following ranges: heights from 0.15 to 2.08 $R_{sun}$, angular widths between 132.6$^\circ$ and 49.8$^\circ$, and inclination angles varying from 24.4 to 5.2$^\circ$ (Columns 1 and 3). Columns 2 and 4 display their corresponding orthogonal projections onto the solar surface. The top-view perspective (Columns 1 and 3) provides optimal visualization of the cone reconstructions, while the face-on view (Columns~2~and~4) illustrates their projected dimming signatures.

The height range in the cone ensemble corresponds to the assumed radial distances at which the CME maintains a magnetic connection to the dimming region in the low corona. As demonstrated in Figures \ref{fig:cone_projection_1} and \ref{fig:cone_projection_2}, the projected cone width exhibits an inverse dependence on height—broader for lower heights and progressively narrower for greater heights—though this variation remains relatively modest. This scaling ensures that the cone projections consistently match the observed spatial extent of the dimming.

\begin{figure}[h]  
	\centering
	\includegraphics[width=0.9\textwidth]{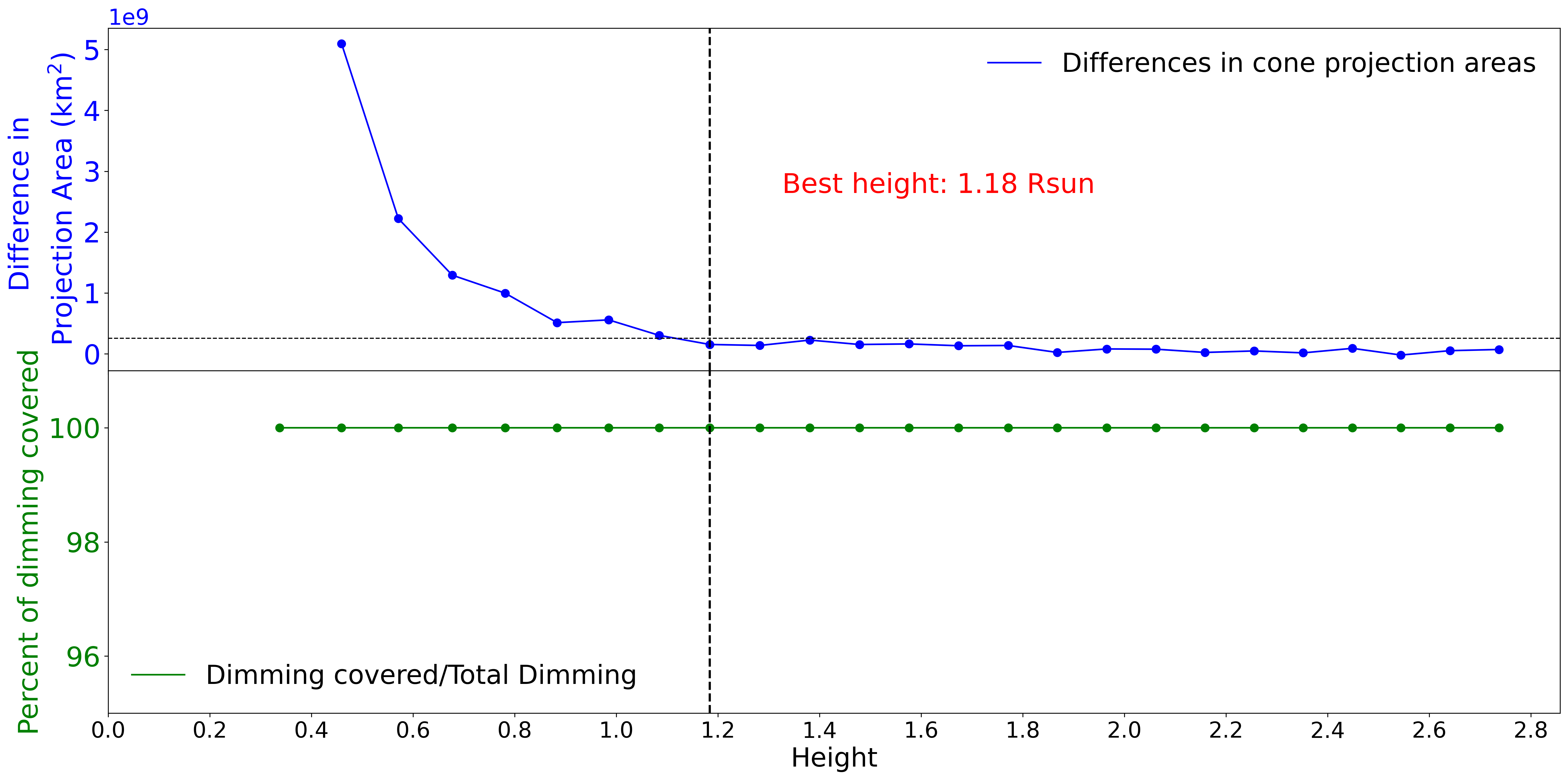}
	\caption{Best-fit criteria for DIRECD cones. The upper panel shows the difference in cone projection areas with 5\% difference shown by dashed black line. The lower shows the dimming coverage for different cone heights} 
	\label{fig:direcd_best_fit_criteria}
\end{figure}

To determine the 3D CME cone configuration that best matches the observed dimming geometry, we implemented a systematic quantification of the cone projection dynamics. For consecutive heights $(h_i, h_{i+1})$, we computed the differential projection area:

\begin{equation}
A_{\Delta h_{i,i+1}} = A_{i+1} - A_i
\end{equation}

where $A_{i+1}$ and $A_i$ represent the projected areas at heights $h_{i+1}$ and $h_i$ respectively. The temporal evolution of $A_{\Delta h_{i,i+1}}$ (Figure \ref{fig:direcd_best_fit_criteria}, top panel, blue curve) reveals a characteristic decrease in projection areas with increasing height. The optimal cone configuration was selected at the point of maximum area contraction rate, where the projection geometry exhibits the highest sensitivity to parameter variations. This critical point was empirically defined at 5\% of the peak differential area (indicated by the vertical dashed line in top panel of Figure~\ref{fig:direcd_best_fit_criteria}). We believe this moment corresponds to the early shrinking stage of the dimming footprint, when the CME is still magnetically connected to the low corona and the on-disk morphology remains sensitive to the cone’s height and angular width. At later times, the footprint becomes nearly stable, and different cone heights produce indistinguishable projections, leading to degeneracy. For the present event \#12, this criterion yielded a CME cone with height 1.18~$R_{sun}$, and corresponding $\beta$ of $21.2~^\circ$, and width of $58^\circ$. In the bottom panel, we see that our ensemble of cones always covers the dimming fully for all heights.

Next, we obtain the best-fit cone inclination within the meridional and equatorial planes. We defined the planes in \citep{jain2024coronal}, where the meridional plane intersects the center O of the Sun, source C on the solar surface, and the North Pole. The equatorial plane passes through point C and is aligned parallel to the equatorial plane of the Sun. In Table \ref{table:direcd_results}, we list the results obtained from DIRECD method for all the events including the best-fit height and inclination angle.

\begin{table}[]
\raggedright
\caption{Results from DIRECD}
\begin{tabular}{|c|c|c|c|c|c|c|c|}

\hline
\# &
\textbf{Event Date} &
\textbf{End Time} &
\textbf{Best Height} &
\textbf{Best Incl. } &
\textbf{Meridional} &
\textbf{Equatorial} \\
  & & (UT) & ($R_{sun}$) & ($^\circ$) & ($^\circ$) & ($^\circ$) \\ \hline

1  & 16-07-2010 & 15:43:00 & 0.55 & 8.2  & $-8.0$  & 1.6   \\ \hline
2  & 01-08-2010 & 08:17:00 & 0.93 & 4.8  & 2.6   & $-4.3$   \\ \hline
3  & 07-08-2010 & 18:39:00 & 1.31 & 24.5 & $-11.7$  & $-21.7$ \\ \hline
4  & 13-02-2011 & 18:33:36 & 0.47 & 8.0  & 5.4    & -6.2  \\ \hline
5  & 15-02-2011 & 02:53:00 & 0.79 & 1.0  & $-0.4$   & 1.0   \\ \hline
6  & 07-03-2011 & 14:58:00 & 0.85 & 14.1 & 7.9   & $-12.5$ \\ \hline
7  & 21-06-2011 & 03:25:36 & 0.59 & 9.7  & $0.0$  & $-10.0$  \\ \hline
8  & 02-08-2011 & 06:56:00 & 0.81 & 9.1   & $-6.8$  & 6.1    \\ \hline
9  & 03-08-2011 & 14:58:00 & 0.71 & 13.8 & $-12.4$ & 6.2   \\ \hline
10 & 06-09-2011 & 02:51:00 & 0.85 & 16.3 & 16.3  & 0.6   \\ \hline
11 & 27-09-2011 & 21:30:29 & 1.1 & 11.2 & $-10.8$ & $-2.8$  \\ \hline
12 & 26-11-2011 & 07:39:00 & 1.18 & 21.2  & $-9.9$  & 18.8  \\ \hline
13 & 25-12-2011 & 19:23:00 & 0.85 & 14.2 & $-5.7$  & 14.7   \\ \hline
14 & 26-12-2011 & 11:53:24 & 0.57 & 4.9   & 3.9    & 3.2   \\ \hline
15 & 26-12-2011 & 02:59:07 & 1.25 & 14.8 & 6.6   & 13.6  \\ \hline
16 & 19-01-2012 & 15:54:00 & 1.05 & 21.0 & 19.5  & $-13.0$ \\ \hline
17 & 07-03-2012 & 00:59:12 & 1.41 & 27.9  & 22.2   & $-22.3$ \\ \hline
18 & 09-03-2012 & 04:07:48 & 0.59 & 3.7  & $-2.7$  & 2.6   \\ \hline
19 & 10-03-2012 & 18:10:48 & 0.6  & 4.9  & 4.8   & 1.0   \\ \hline
20 & 05-04-2012 & 21:18:00 & 0.63 & 7.2  & 3.9   & 6.8   \\ \hline
21 & 03-06-2012 & 18:28:00 & 0.99 & 18.3 & 11.0  & $-16.5$  \\ \hline
22 & 06-06-2012 & 20:47:48 & 1.05 & 17.3 & $-17.2$ & $-1.4$   \\ \hline
23 & 14-06-2012 & 14:39:36 & 0.59 & 9.2  & $-5.6$  & $-7.9$  \\ \hline
24 & 04-07-2012 & 17:26:23 & 0.72 & 3.6   & $-2.5$  & 2.7   \\ \hline
25 & 25-09-2012 & 04:48:35 & 1.12 & 2.0  & $-1.0$  & 1.8    \\ \hline
26 & 11-04-2013 & 08:24:11 & 0.86 & 9.5  & $-5.5$  & $-7.8$  \\ \hline
27 & 17-05-2013 & 10:06:23 & 0.94 & 17.3 & 13.9   & $-11.5$ \\ \hline
28 & 06-10-2013 & 14:41:11 & 0.62 & 16.0 & 11.6  & $-11.1$ \\ \hline
29 & 10-11-2013 & 05:42:59 & 0.74 & 13.8 & $-12.4$ & $-6.9$  \\ \hline
30 & 18-04-2014 & 07:54:59 & 0.91 & 19.8 & $-19.6$ & 4.2    \\ \hline
31 & 20-12-2014 & 01:34:59 & 1.36 & 35.2 & $-27.3$ & 34.5   \\ \hline
32 & 08-03-2019 & 03:33:57 & 0.52 & 1.1  & 0.7   & 0.8   \\ \hline
33 & 28-10-2021 & 16:22:09 & 0.46 & 11.7 & $-7.4$  & $-11.1$ \\ \hline
\end{tabular}%
\tablecomments{We list here the event date, the end of the impulsive phase, the height of the best-fit cone (in $R_{sun}$) obtained from DIRECD, the inclination angle of the best-fit cone in 3D and the inclination angles of the best-fit cone in meridional/equatorial planes.}

\label{table:direcd_results}
\end{table}

\section{Results}

To validate our results, we performed comparative analyses with both GCS reconstructions of the events and visual comparisons with SOHO/LASCO coronagraph observations \citep{thompson1998soho}. For this validation, we utilized the comprehensive LLAMACoRe catalog \citep{kay2024collection}, which provides GCS reconstructions for 28 out of our 33 selected events.
For the two events of 2019 and 2021, we directly take GCS reconstructions from \citep{dumbovic20212019} and \citep{Chikunova2023} giving us reliable GCS reconstructions for 30 out of 33 events analyzed. From these reconstructions, we extracted the GCS latitude and longitude parameters, representing the central coordinates of the CME bubble on the solar sphere. We focused our validation on two primary outputs from DIRECD: the 3D inclination direction and the inclination angles in planes. 

For the quantitative comparison of the 3D DIRECD inclination angle with GCS results, we computed the 3D angular separation between the vector connecting the solar center to the flare source location and the vector connecting the solar center to the GCS-derived latitude/longitude coordinates and define it as the inclination angle from GCS. These angles with respect to the radial direction as estimated from the location of the associated flare, give an estimate of the latitudinal and longitudinal inclination of the CME from the radial propagation direction in 3D. They were then compared with the inclination angles derived from DIRECD. A critical consideration in this comparison arises from the different coordinate systems employed - while GCS uses a Sun-centered reference frame, DIRECD's angles are computed from a surface-centered perspective at the flare location. This difference in reference frames introduces a systematic offset in angle measurements. However, we note that this discrepancy diminishes with increasing CME propagation distance, as the angular separation between surface-centered and origin-centered vectors converges for distant features.

To compare the CME direction results obtained from DIRECD and GCS, we first calculate the latitude and longitude differences between the flare location and GCS coordinates on the Sun's surface. These differences effectively represent the 2D angular separation separated into two components. We then directly compare the DIRECD's meridional plane angles with the latitude differences, and DIRECD's equatorial plane angles with the longitude differences. This approach provides a straightforward way to validate DIRECD's measurements against GCS in 2D planes. In Table \ref{table:GCS_DIRECD_comparison}, we list the 3D inclination angle derived from GCS, best inclination angle from DIRECD in 3D, latitude and longitude of the GCS source coordinates, latitude and longitude of DIRECD, which coincides with the flare source coordinates, the latitudinal and longitudinal inclinations from the radial propagation (lat/lon difference) and angles in Meridional and Equatorial planes derived from DIRECD.

\begin{table}[]
\movetabledown=5.5cm
\begin{rotatetable}

\caption{Comparison of GCS and DIRECD}
\raggedright
{\scriptsize

\begin{tabular}{|c|c|c|c|c|c|c|c|c|c|c|c|c|}
\hline
\# &
  \textbf{Event Date} &
  \textbf{Incl. in 3D} &
  \textbf{Incl. in 3D} &
  \textbf{Source Lat.} &
  \textbf{Source Lat.} &
  \textbf{Lat. Difference} &
  \textbf{Meridional} &
  \textbf{Source Lon.} &
  \textbf{Source Lon.} &
  \textbf{Lon. Difference} &
  \textbf{Equatorial} \\
  & & \textbf{(GCS)} &\textbf{(DIRECD)} & \textbf{(GCS)} & \textbf{(DIRECD)} &\textbf{(GCS - DIRECD) } & \textbf{(DIRECD)} & \textbf{(GCS) } & \textbf{(DIRECD)}&\textbf{(GCS - DIRECD)}  & \textbf{(DIRECD)}\\\hline
1  & 16-07-2010 & ...    & 8.2  & ...   & $-21$ & ...    & $-8.1$  & ...   & 20  & ...    & 1.6   \\ \hline
2  & 01-08-2010 & 6.1  & 4.8  & 18 & 20  & $-2$    & 2.6   & -28.9  & $-35$ & 6.1   & $-4.3$   \\ \hline
3  & 07-08-2010 & 16.7 & 24.5 & $-2.1$   & 14  & $-16.1$ & $-11.7$  & $-41.5$    & $-37$ & $-4.5$  & $-21.7$ \\ \hline
4  & 13-02-2011 & 10.9 & 8.0  & $-10.5$ & $-20$ & 9.5   & 5.4    & $-0.5$  & $-5$  & 4.5   & $-6.2$  \\ \hline
5  & 15-02-2011 & 12.1 & 1.0  & $-10$   & $-20$ & 10    & $-0.4$   & 5    & 12  & $-7$    & 1.0   \\ \hline
6  & 07-03-2011 & 6.3  & 14.1 & 16   & 12  & 4     & 7.9   & $-16$    & $-21$ & 5     & $-12.5$ \\ \hline
7  & 21-06-2011 & 18.3 & 9.7  & 7 & 14  & $-7$    & 0.0  & $-7.2$  & 10  & $-17.2$ & $-9.9$  \\ \hline
8  & 02-08-2011 & 4.0  & 9.1   & 11.2 & 15  & $-3.8$  & $-6.8$  & 15.3    & 14  & 1.3   & 6.1    \\ \hline
9  & 03-08-2011 & 19.3 & 13.8 & 15.5 & 17  & $-1.5$  & $-12.4$ & 10      & 30  & $-20$   & 6.2   \\ \hline
10 & 06-09-2011 & 15.0 & 16.3 & 28.6 & 14  & 14.6  & 16.3  & 10.5     & 7   & 3.5   & 0.6   \\ \hline
11 & 27-09-2011 & ...    & 11.2 & ...   & 10  & ...    & $-10.8$ & ...       & $-9$  & ...    & $-2.8$  \\ \hline
12 & 26-11-2011 & 13.1 & 21.2  & 17   & 11  & 6     & $-9.9$  & 35       & 47  & $-12$   & 18.8  \\ \hline
13 & 25-12-2011 & 19.4 & 14.2 & -8   & $-22$ & 14    & $-5.7$  & 40       & 26  & 14    & 14.7   \\ \hline
14 & 26-12-2011 & 3.1  & 4.9   & 22.9  & 20  & 2.9   & 3.9    & 0       & 1   & $-1$    & 3.2   \\ \hline
15 & 26-12-2011 & 13.3 & 14.8 & $-19$  & $-18$ & $-1$    & 6.6   & 44       & 30  & 14    & 13.6  \\ \hline
16 & 19-01-2012 & 13.2 & 21.0 & 44   & 32  & 12    & 19.5  & $-20.1$   & $-27$ & 6.9   & $-13.0$ \\ \hline
17 & 07-03-2012 & 15.2 & 27.9  & 33  & 18  & 15    & 22.2   & $-33.5$   & $-31$ & -2.5  & $-22.3$ \\ \hline
18 & 09-03-2012 & 11.5 & 3.7  & 5.5 & 17  & $-11.5$ & $-2.7$  & 1       & $-2$  & 3     & 2.6   \\ \hline
19 & 10-03-2012 & 12.6 & 4.9  & 18   & 16  & 2     & 4.8   & 11  & 24  & $-13$   & 1.0   \\ \hline
20 & 05-04-2012 & 40.2 & 7.2  & 26   & 24  & 2     & 3.9   & $-12.5$ & 32  & $-44.5$ & 6.8   \\ \hline
21 & 03-06-2012 & 18.8 & 18.3 & 33   & 17  & 16    & 11.04  & $-49$  & $-38$ & $-11$   & $-16.5$  \\ \hline
22 & 06-06-2012 & 15.6 & 17.3 & $-29.1$ & $-18$ & $-11.1$ & $-17.2$ & $-7$    & 5   & $-12$   & $-1.4$   \\ \hline
23 & 14-06-2012 & 6.9  & 9.2  & $-22.4$   & $-17$ & $-5.4$  & $-5.6$  & $-0.4$   & $-5$  & 4.6   & $-7.9$  \\ \hline
24 & 04-07-2012 & 35.3 & 3.6   & 5 & 12  & $-7$    & $-2.5$  & 0   & 35  & $-35$   & 2.7   \\ \hline
25 & 25-09-2012 & ...    & 2.0  & ...  & 9   & ...    & $-1.0$  & ...      & $-20$ & ...    & 1.8    \\ \hline
26 & 11-04-2013 & 10.1 & 9.5  & $-1$   & 9   & $-10$   & $-5.5$  & $-11$    & $-12$ & 1     & $-7.8$  \\ \hline
27 & 17-05-2013 & 9.8  & 17.3 & 13    & 12  & 1     & 13.9   & $-32$      & $-42$ & 10    & $-11.5$ \\ \hline
28 & 06-10-2013 & 11.4 & 16.0 & $-6.1$  & $-15$ & 8.9   & 12.0  & 4.7     & 12  & $-7.3$  & $-11.1$ \\ \hline
29 & 10-11-2013 & 18. & 13.8 & $-31$   & $-14$ & $-17$   & $-12.4$ & 22     & 14  & 8     & $-6.9$  \\ \hline
30 & 18-04-2014 & 24.1 & 19.8 & $-28.8$ & $-23$ & $-5.8$  & $-19.6$ & 12.9    & 39  & $-26.1$ & 4.2    \\ \hline
31 & 20-12-2014 & 22.0 & 35.2 & $-43$  & $-21$ & $-22$   & $-27.3$ & 23       & 24  & $-1$    & 34.5   \\ \hline
32 & 08-03-2019 & 3.6   & 1.1  & 6   & 9   & $-3$    & 0.7   & 5        & 3   & 2     & 0.8   \\ \hline
33 & 28-10-2021 & 4.0  & 11.7 & $-30$  & $-28$ & $-2$    & $-7.4$  & $-3$      & 1   & $-4$    & $-11.1$ \\ \hline
\end{tabular}%
\tablecomments{We list here the inclination angle from GCS in 3D, best inclination angle from DIRECD in 3D, Latitude/Longitude of GCS source coordinates, Latitude/Longitude of DIRECD source coordinates, the latitudinal and longitudinal inclinations from the radial propagation (Lat/lon difference) between GCS and DIRECD, and the angles in Meridional and Equatorial planes derived from DIRECD for all events. All values are in degrees.}

\label{table:GCS_DIRECD_comparison}
}

\end{rotatetable}
\end{table}

\begin{figure}[h]
\centering 
\subfloat[]{%
  \includegraphics[width=0.3\columnwidth]{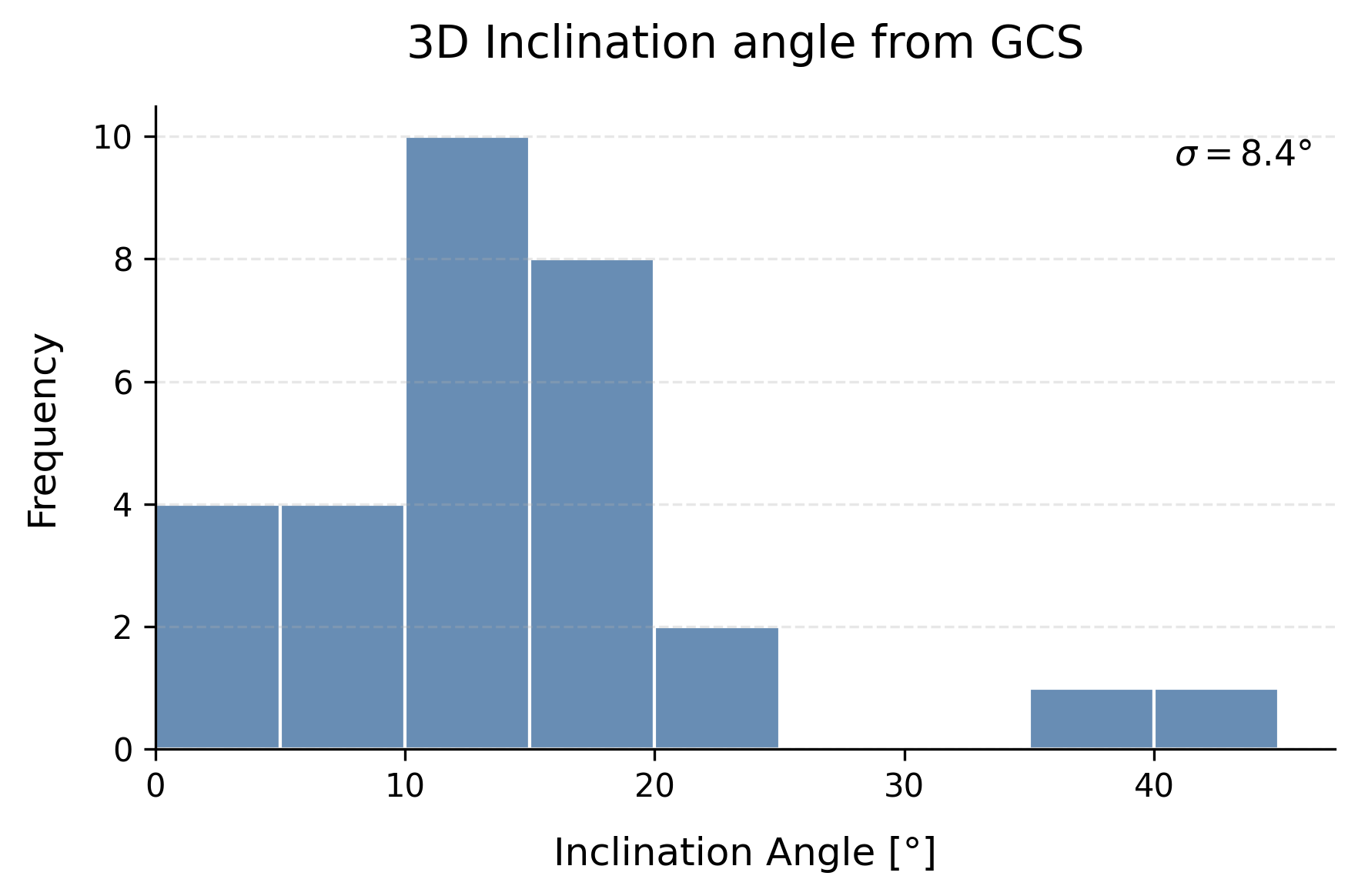}%
  \label{plot:histogram_gcs_source_3D}%
}\qquad
\subfloat[]{%
  \includegraphics[width=0.3\columnwidth]{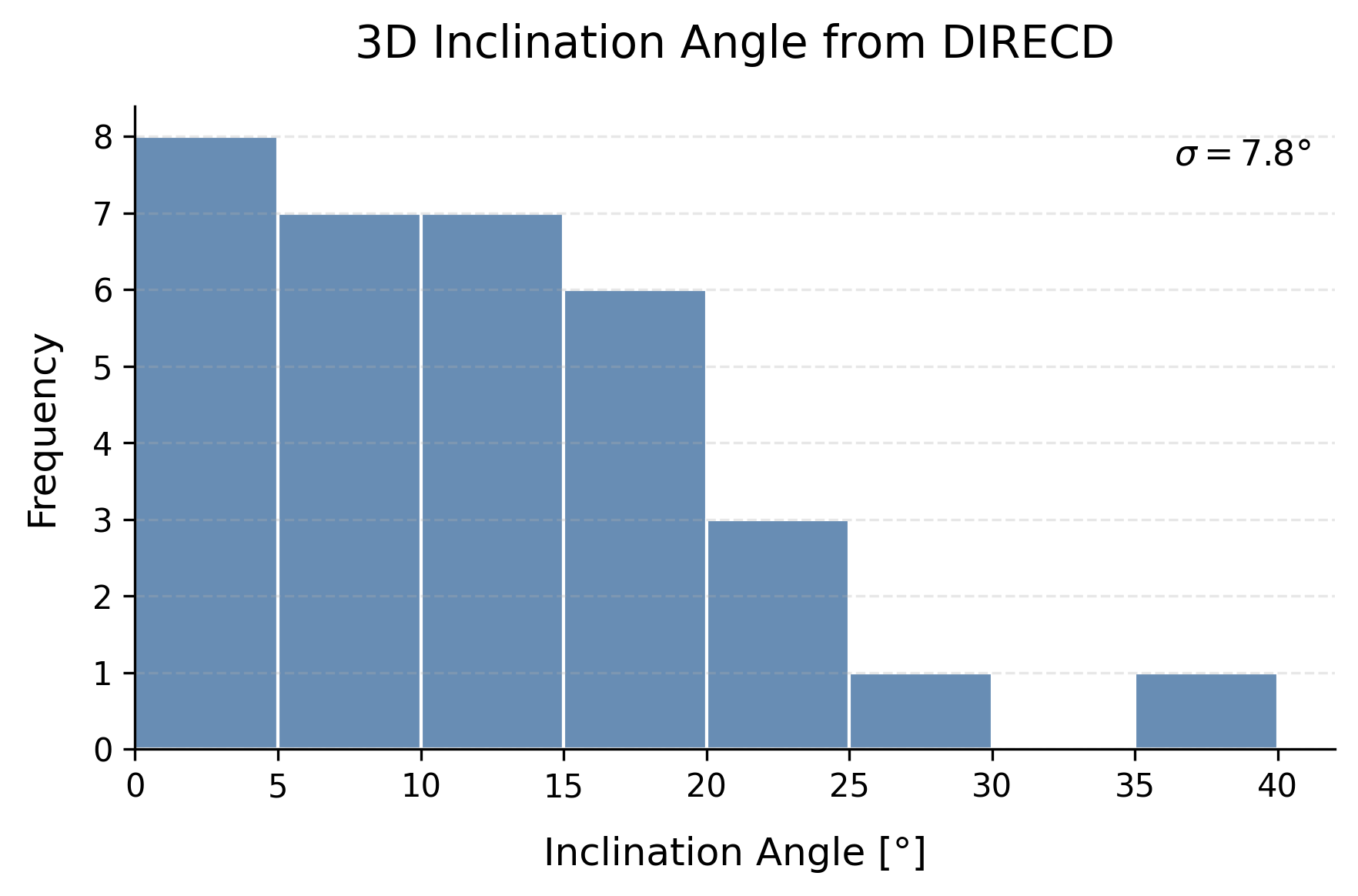}%
  \label{plot:histogram_inclination}%
}\qquad
\subfloat[]{%
  \includegraphics[width=0.3\columnwidth]{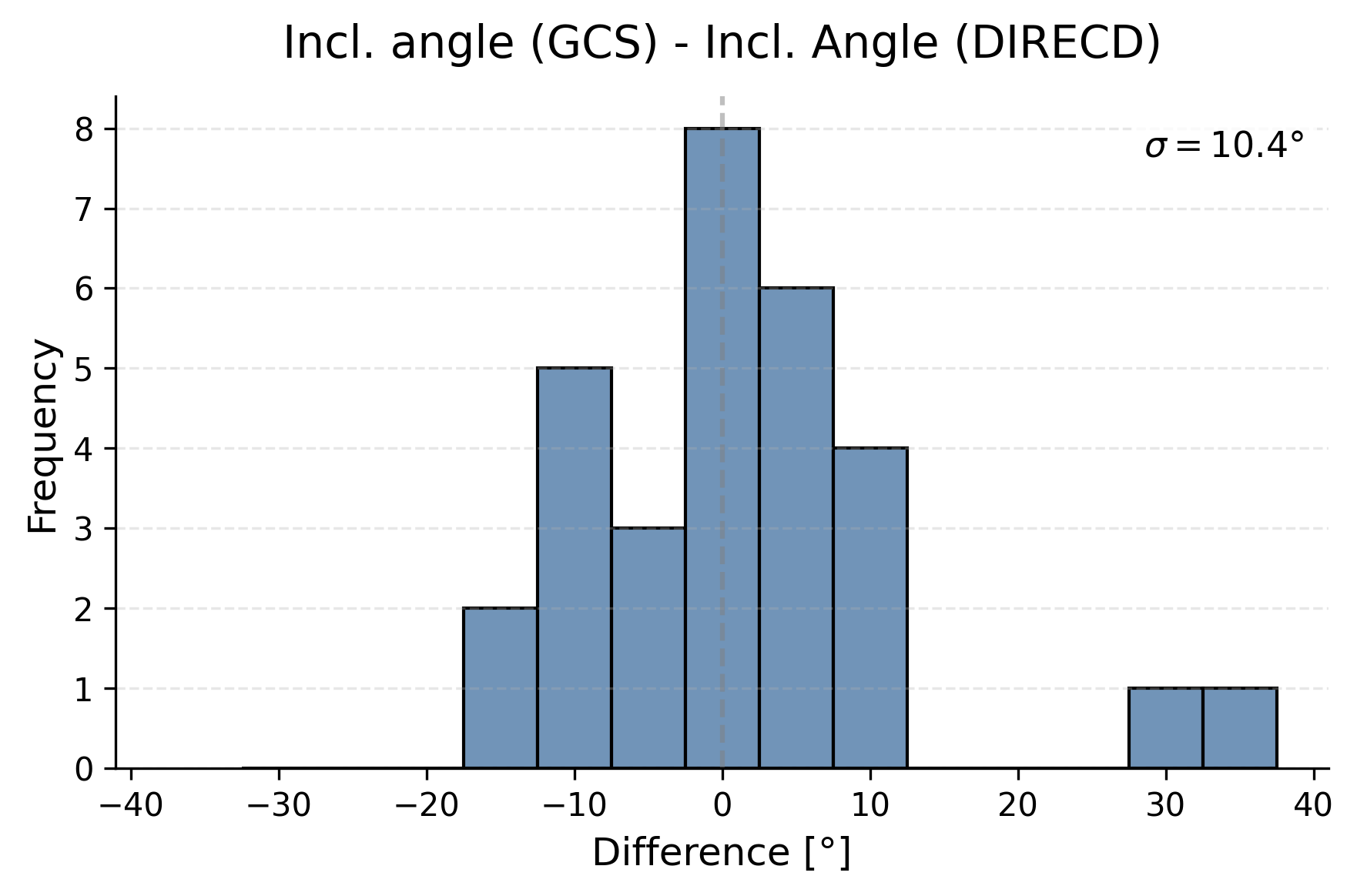}%
  \label{plot:histogram_gcs_direcd_3D}%
}

\caption{Left Panel: Histogram of the 3D inclination angle derived from GCS. Middle Panel: Histogram of the 3D inclination angle derived from DIRECD. Right Panel: Histogram of the differences in inclination angles between GCS and DIRECD.}
\label{histogram_gcs_direcd}
\end{figure}

Figure \ref{histogram_gcs_direcd} presents the 3D inclination angles between CME and radial direction derived from GCS (panel a), DIRECD-derived inclination angles (panel b), and their differences (panel c), respectively. Our analysis reveals a general agreement between the results from the GCS and DIRECD methods, with a mean absolute error $1.2^\circ$ and a standard deviation of differences $\sigma_d = 10.4^\circ$. The Anderson-Darling test results (A-D statistic = 0.569, p = 0.193) indicate no statistically significant difference between the two datasets.

\begin{figure}
\centering 
\subfloat[]{%
  \includegraphics[width=0.3\columnwidth]{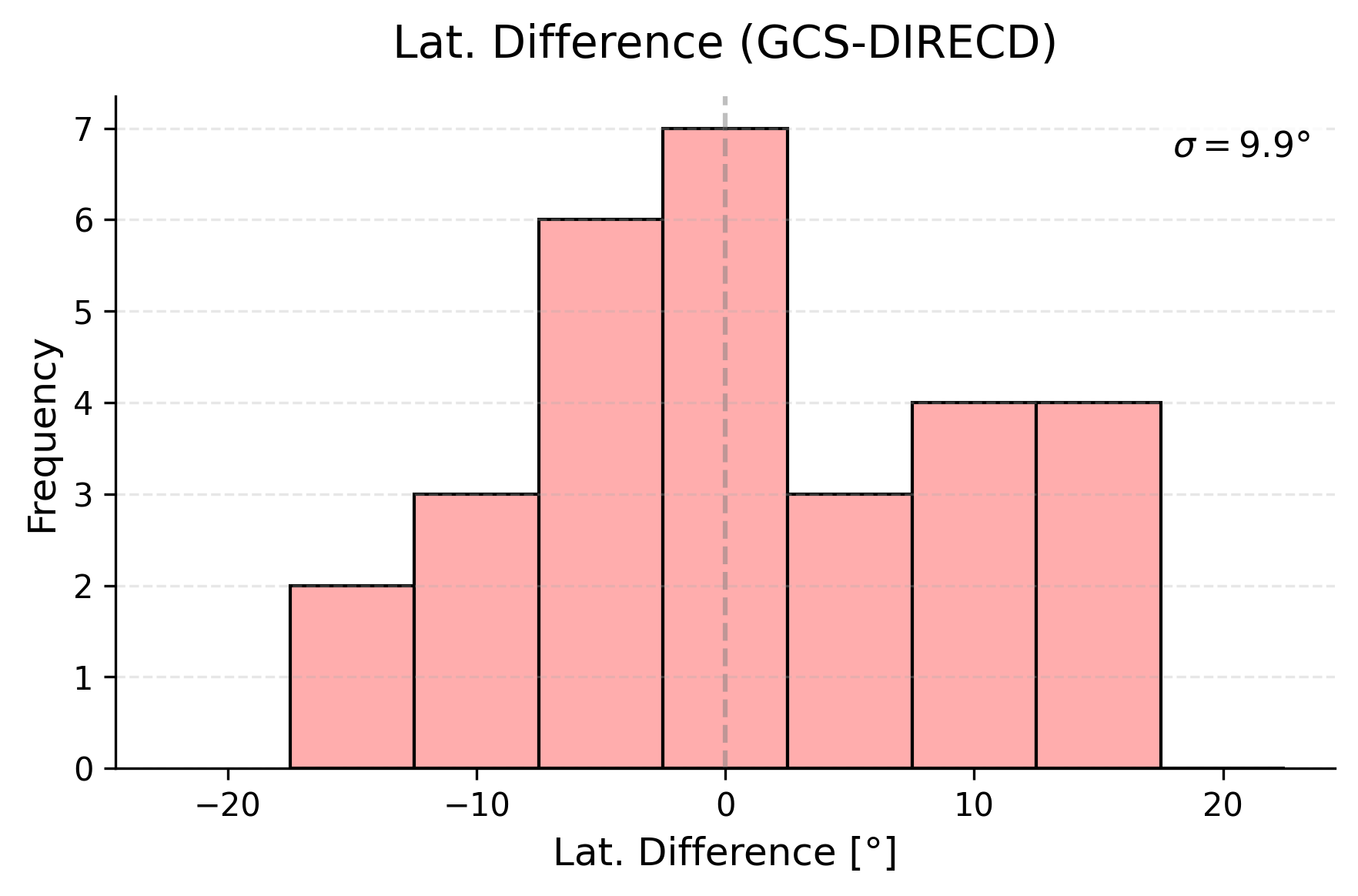}%
  \label{plot:histogram_gcs_source_lat}%
}\qquad
\subfloat[]{%
  \includegraphics[width=0.3\columnwidth]{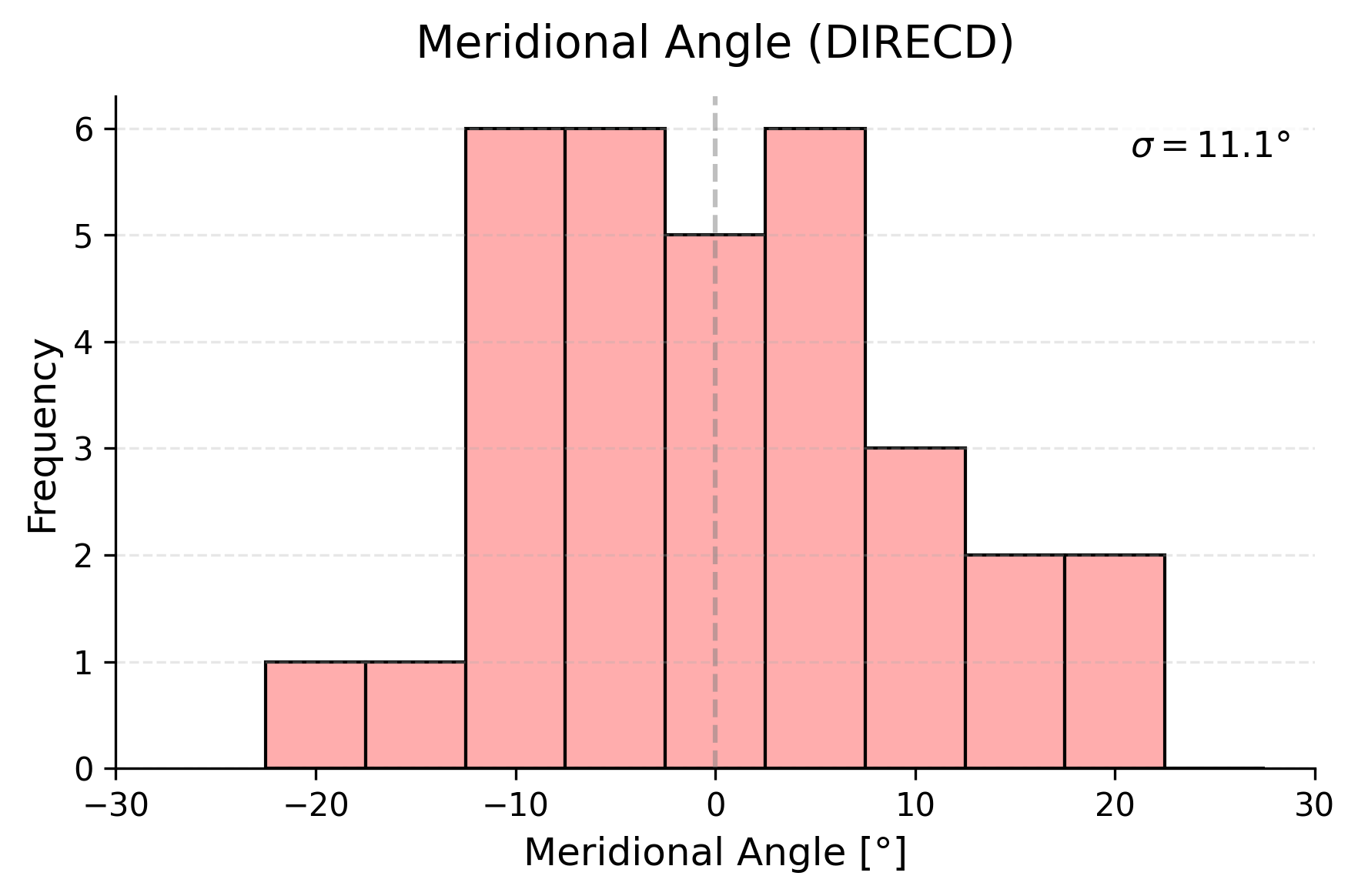}%
  \label{plot:histogram_direcd_source_lat}%
}\qquad
\subfloat[]{%
  \includegraphics[width=0.3\columnwidth]{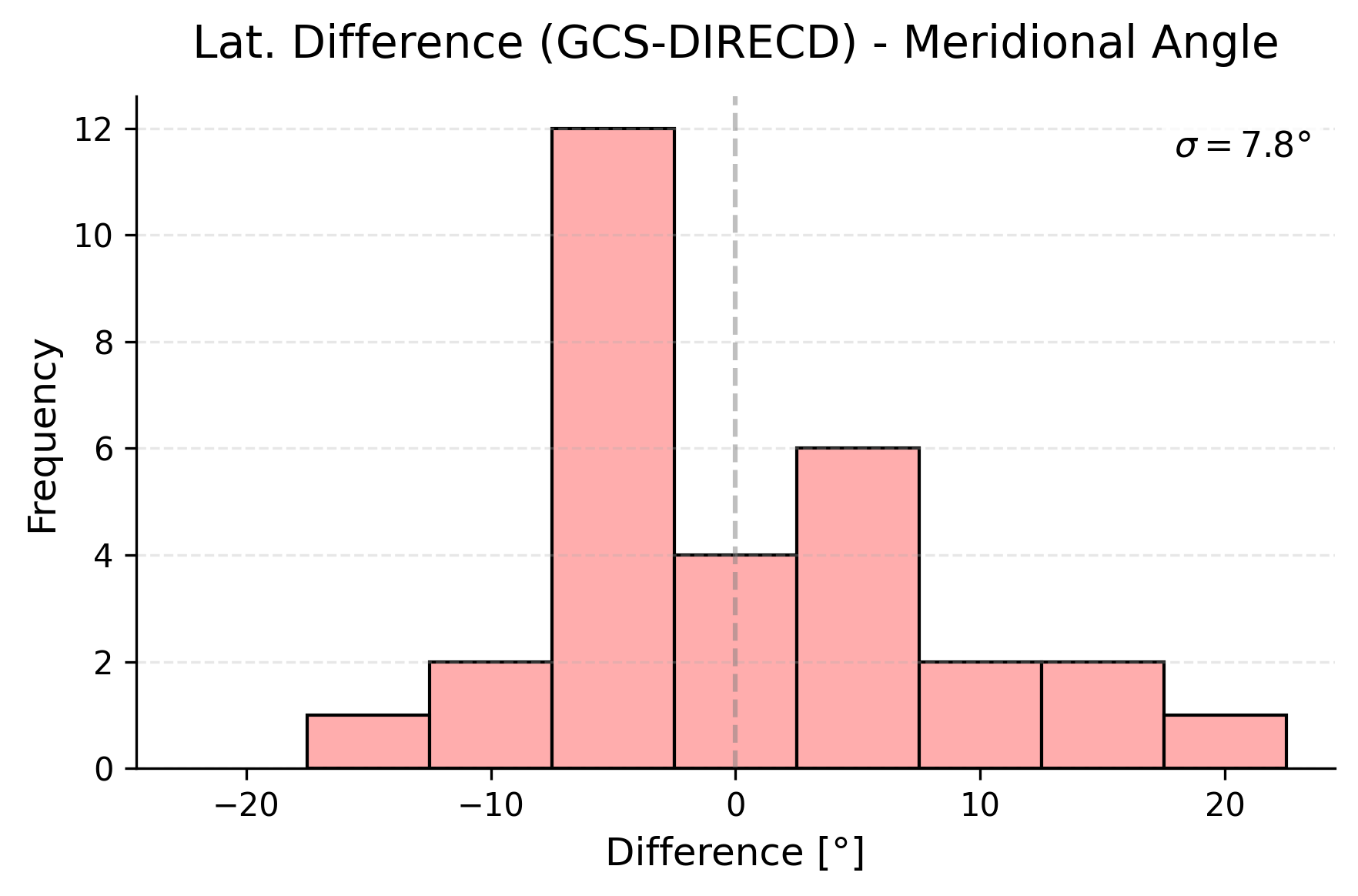}%
  \label{plot:histogram_gcs_direcd_lat}%
}

\caption{Left panel: Histogram of latitude difference between GCS and DIRECD source locations. Middle panel: Histogram of meridional angle from DIRECD. Right Panel: Histogram of differences of the latitudinal inclinations of the CME direction from radial as derived from GCS and DIRECD}
\end{figure}

We further validate DIRECD by comparing its meridional and equatorial angles to latitudinal and longitudinal differences between DIRECD and GCS source. Figure \ref{plot:histogram_gcs_source_lat} displays the latitude differences between GCS and DIRECD source, while Figure \ref{plot:histogram_direcd_source_lat} shows the meridional angles from DIRECD. The histogram of the differences of the latitudinal inclinations of the CME direction from radial as derived from GCS and DIRECD shown in Figure \ref{plot:histogram_gcs_direcd_lat} demonstrates a very good agreement between the two methods, with a mean absolute error of $0.3^\circ$ and a standard deviation of $7.8^\circ$. The Anderson-Darling test (high p-value of 0.95) strongly suggest the two datasets are close to each other.

\begin{figure}
\centering 
\subfloat[]{%
  \includegraphics[width=0.3\columnwidth]{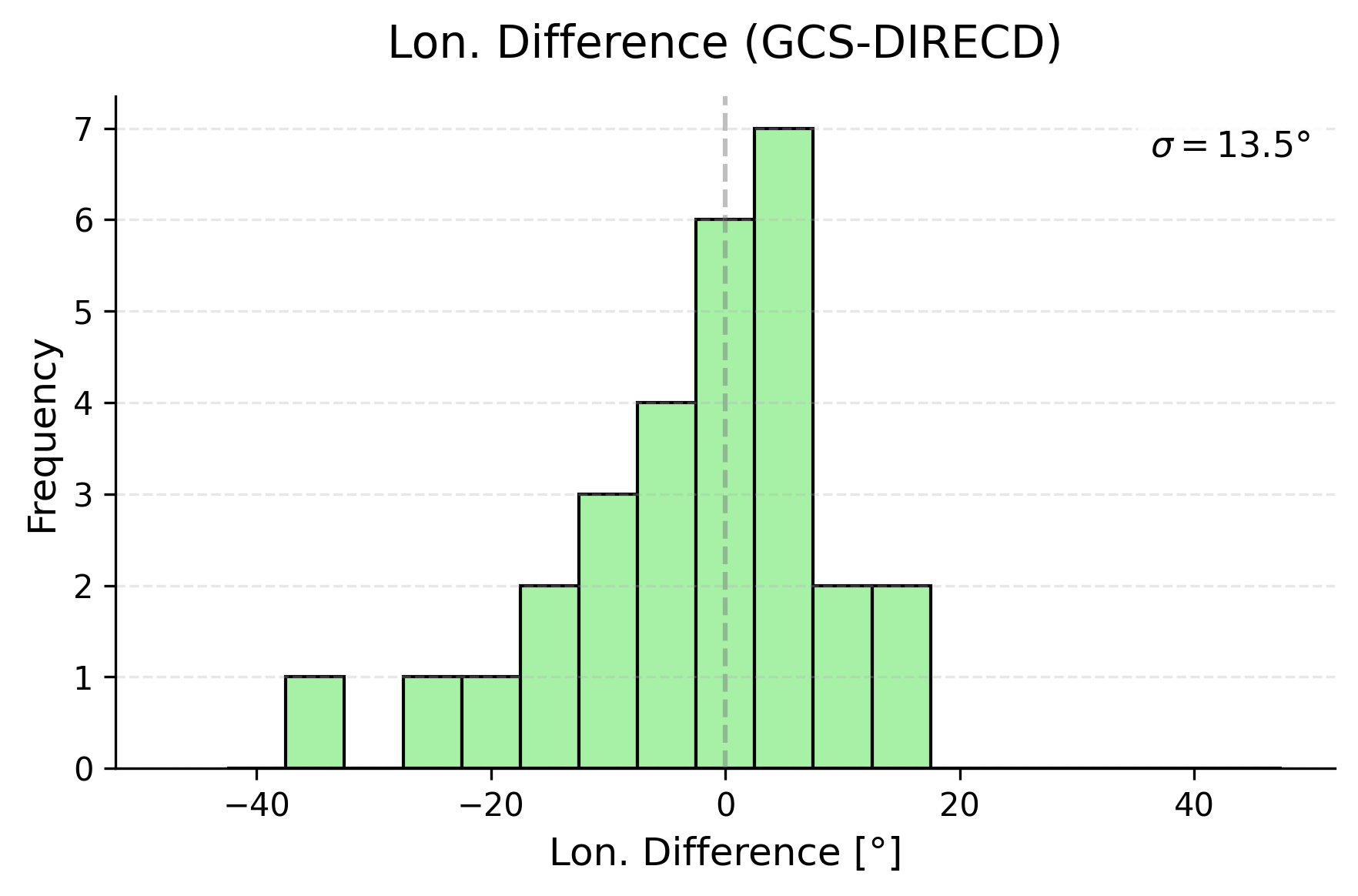}%
  \label{plot:histogram_gcs_source_lon}%
}\qquad
\subfloat[]{%
  \includegraphics[width=0.3\columnwidth]{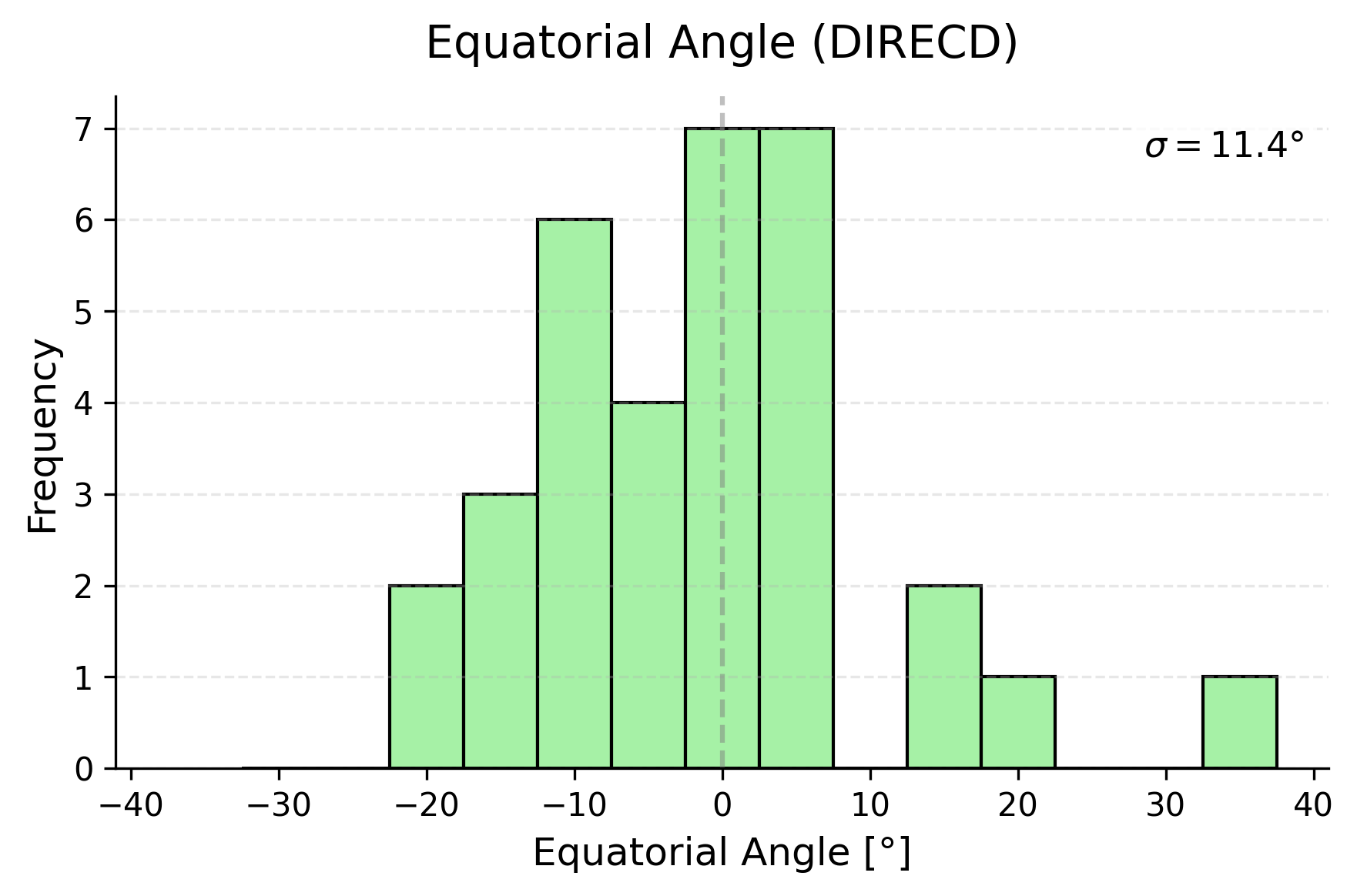}%
  \label{plot:histogram_direcd_source_lon}%
}\qquad
\subfloat[]{%
  \includegraphics[width=0.3\columnwidth]{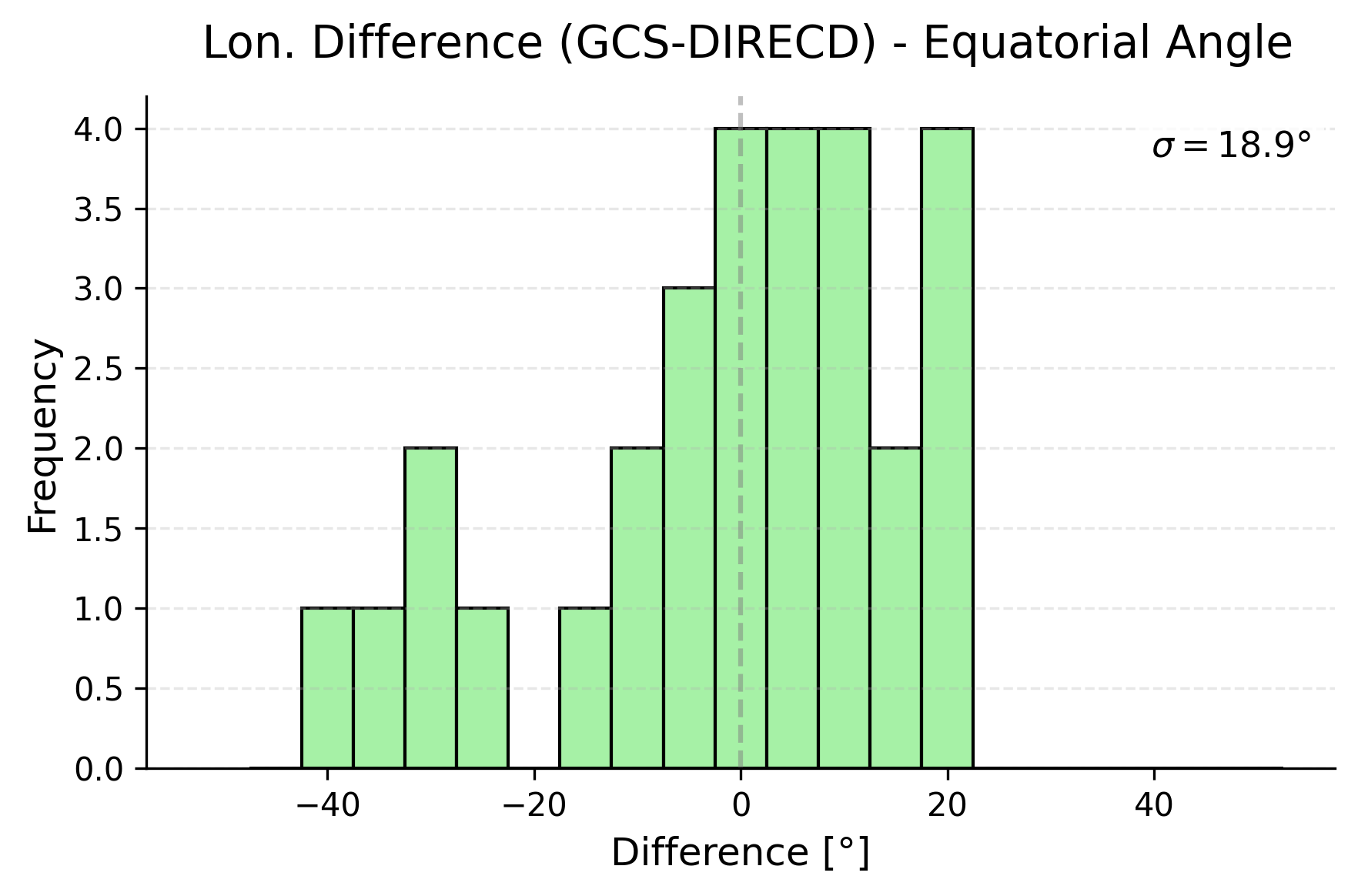}%
  \label{plot:histogram_gcs_direcd_lon}%
}

\caption{Left panel: Histogram of longitude difference between GCS and DIRECD. Middle panel: Histogram of equatorial angle from DIRECD. Right Panel: Histogram of differences of the longitudinal inclinations of the CME direction from radial as derived from GCS and DIRECD.}
\end{figure}

Figure \ref{plot:histogram_gcs_source_lon} presents the longitudinal differences between the GCS and DIRECD sources, while Figure \ref{plot:histogram_direcd_source_lon} shows the equatorial plane angles derived from DIRECD. Histogram of differences of the longitudinal inclinations of the CME direction from radial as derived from GCS and DIRECD in Figure \ref{plot:histogram_gcs_direcd_lon} reveals a significant standard deviation ($\sigma_{d} = 18.9^\circ$) and mean absolute error of $-2.9^\circ$, with the GCS source longitude (and thus the longitudinal difference between GCS and DIRECD source) systematically shifted eastward, supported by a significant left-skewed distribution (Fisher-Pearson skewness coefficient = $-1.2$, Galton skewness coefficient = $-0.3$).

\begin{figure}
\centering 
\subfloat[]{%
  \includegraphics[width=0.3\columnwidth]{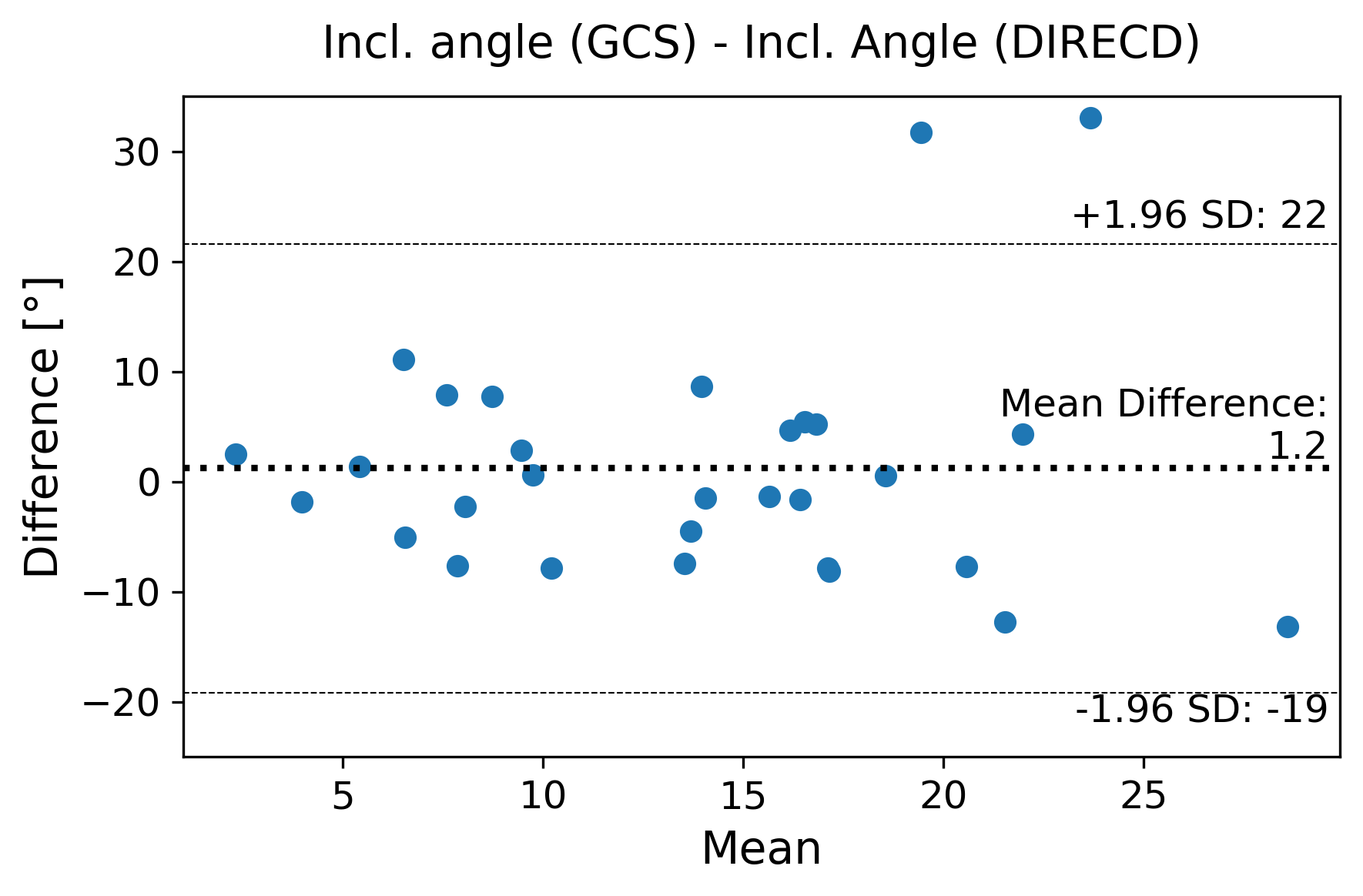}%
  \label{plot:band_altman_gcs_direcd_3d}%
}\qquad
\subfloat[]{%
  \includegraphics[width=0.3\columnwidth]{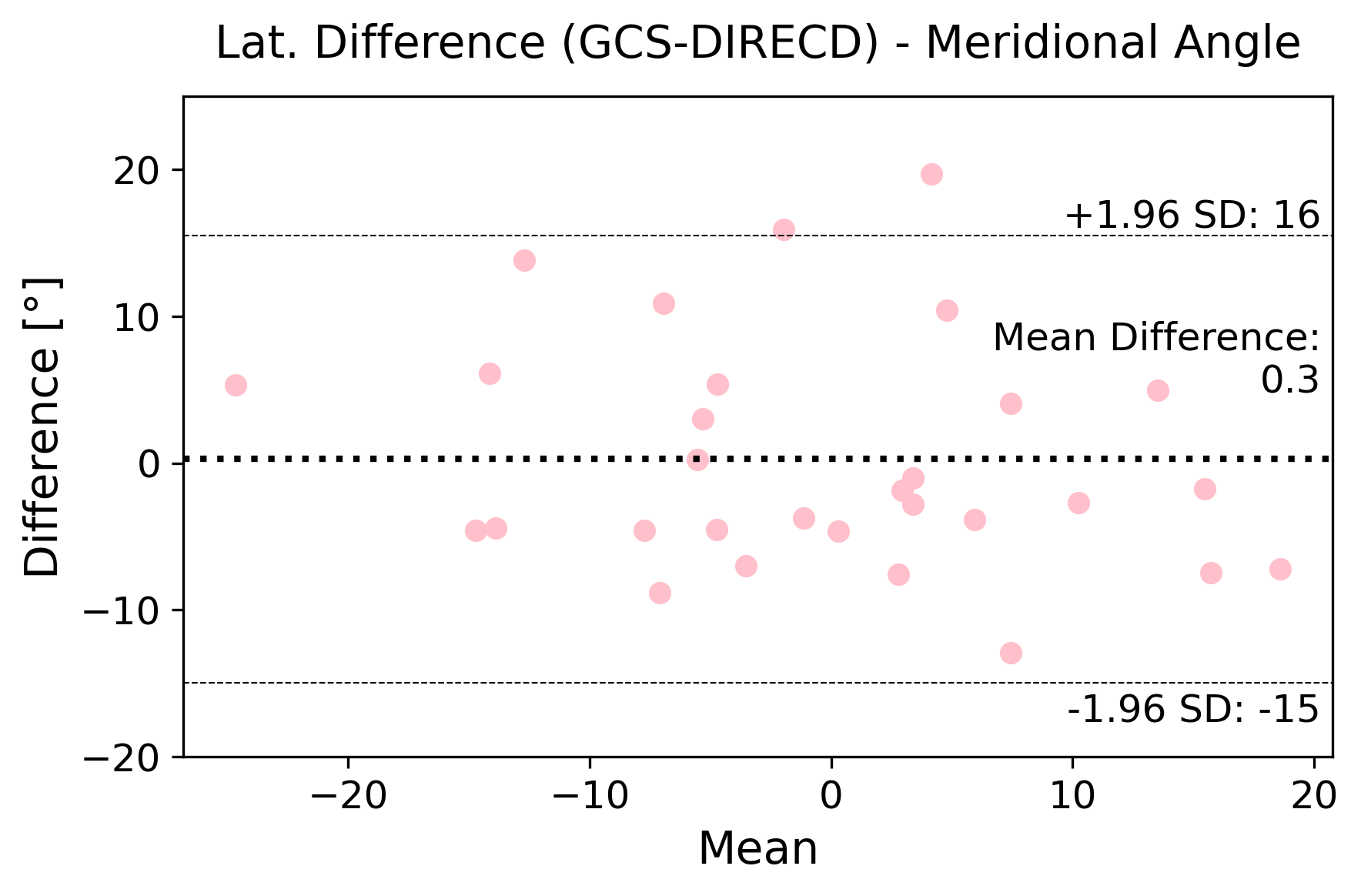}%
  \label{plot:band_altman_gcs_direcd_lat}%
}\qquad
\subfloat[]{%
  \includegraphics[width=0.3\columnwidth]{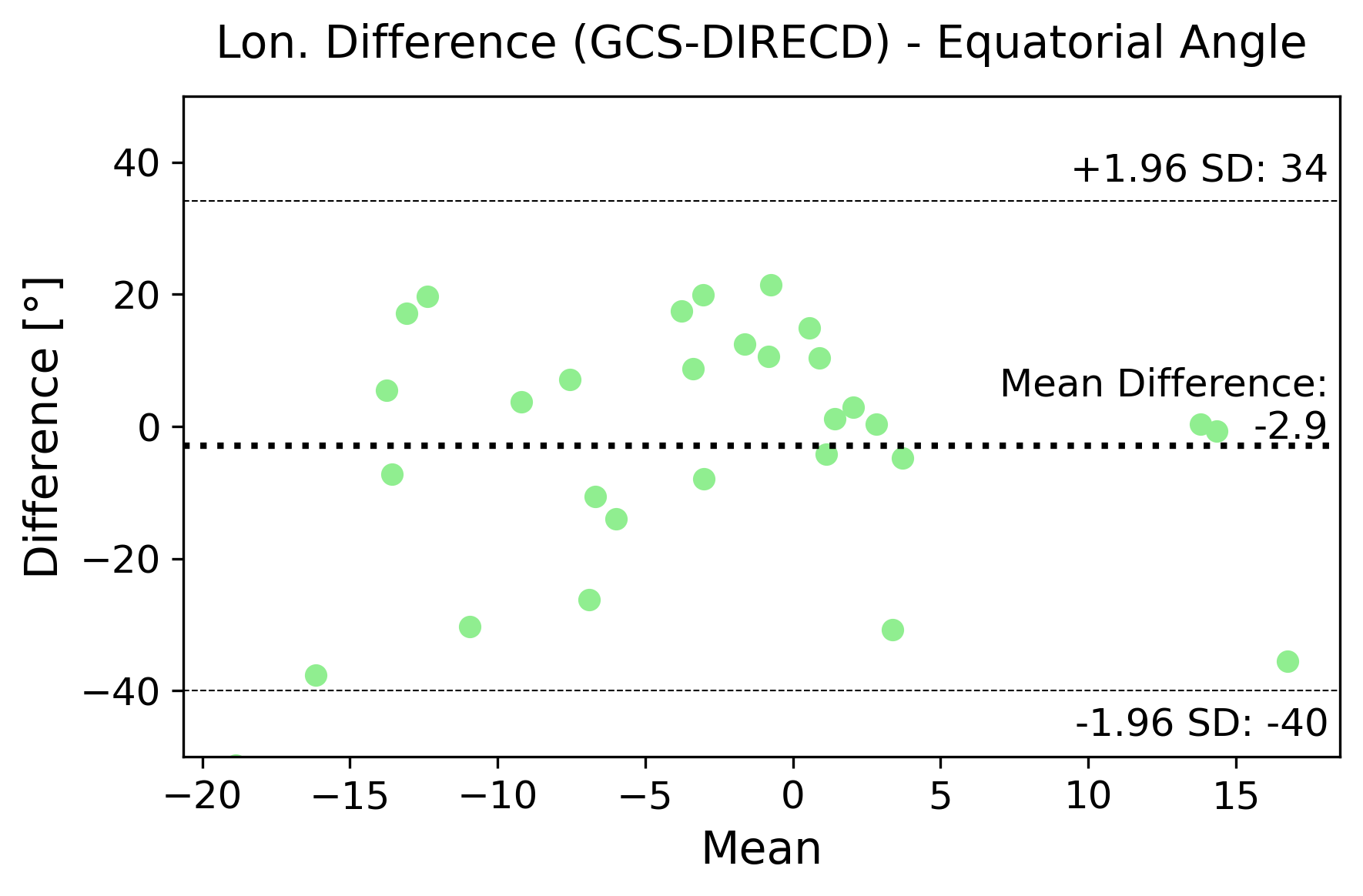}%
  \label{plot:band_altman_gcs_direcd_lon}%
}

\caption{Bland-Altman plot of GCS and DIRECD method for (a) Inclination angle in 3D, (b) Differences of the latitudinal inclinations of the CME direction from radial as derived from GCS and DIRECD, (c) Differences of the longitudinal inclinations of the CME direction from radial as derived from GCS and DIRECD}
\end{figure}

The Bland-Altman analysis demonstrates strong agreement between DIRECD and GCS in determining 3D inclination angles, as illustrated in Figure~\ref{plot:band_altman_gcs_direcd_3d}. The mean difference between the two methods is minimal ($1.2^\circ$), with 95\% of the differences falling within $\pm 22^\circ$. Importantly, no systematic trends or significant bias were detected across the full range of angles.

Further analysis of latitudinal and longitudinal inclinations (Figures~\ref{plot:band_altman_gcs_direcd_lat} and~\ref{plot:band_altman_gcs_direcd_lon}) reveals a negligible mean difference of $0.3^\circ$ between DIRECD and GCS latitudinal inclinations, confirming high consistency. However, the longitudinal inclinations derived from the two methods show a slightly larger mean difference of $-2.9^\circ$, with 95\% of differences within $\pm 40^\circ$. The increased variability here likely stems from Parker spiral effects during later-stage CME propagation, as supported by prior studies \citep{martinic2022determination, martinic2023effects, fargette2022preferential} that influence GCS at larger heliocentric distances, whereas  DIRECD focuses on the initial eruption phase (up to $3R_{\text{sun}}$), where CME direction is least affected by interplanetary dynamics.

\begin{figure}
\centering 
\subfloat[]{%
  \includegraphics[width=0.35\columnwidth]{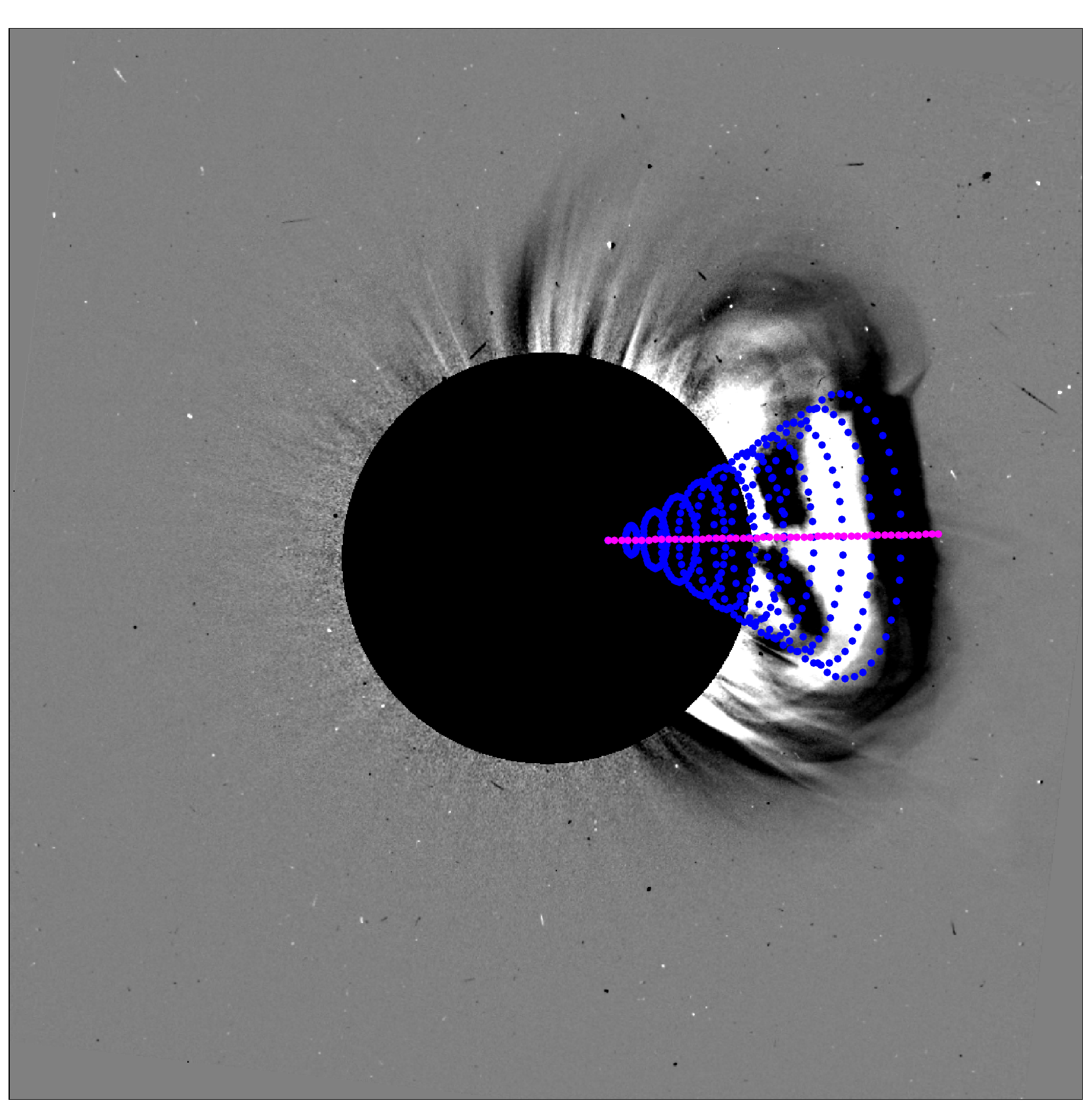}%
  \label{plot:lasco_261111}%
}\qquad
\subfloat[]{%
  \includegraphics[width=0.35\columnwidth]{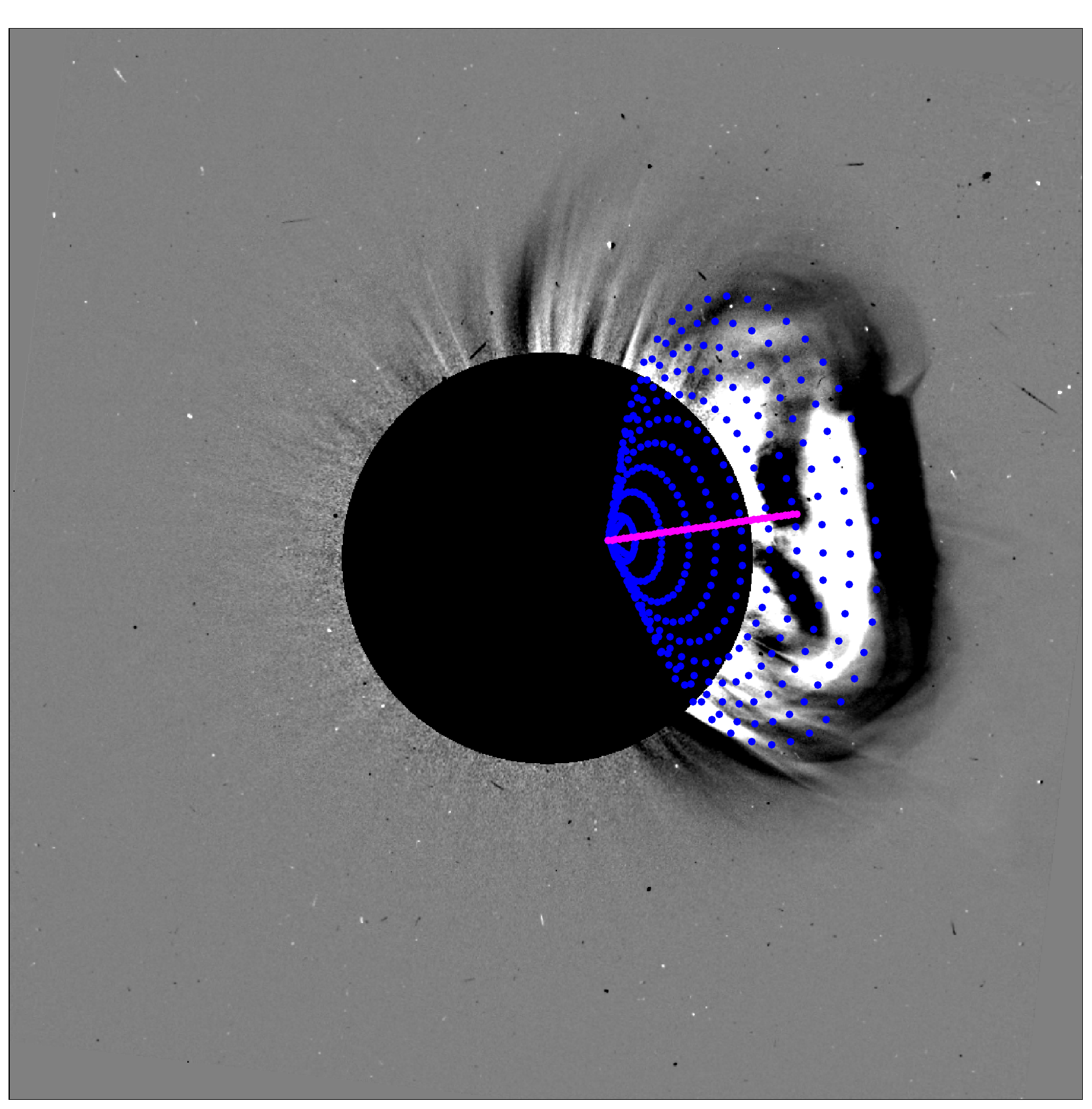}%
  \label{plot:lasco_261111_extended}%

}

\caption{Best-fit DIRECD cones (blue) together with the central cone axis (in pink) superimposed on LASCO coronagraph data for the event on November 26, 2011 at 07:48 UT. The left panel shows the best-fit cone fitted to the dimming at the end of the impulsive phase, while the right panel shows the best-fit cone taking secondary dimmings regions into account.}
\end{figure}

To further validate our DIRECD results with observed CME structures, we projected the best-fit cone onto LASCO/C2 coronagraph images near the end of the impulsive phase. Figure \ref{plot:lasco_261111} presents the best-fit cone model for Event \#12, overlaid on the LASCO/C2 observation (07:48 UT), with the cone’s central axis indicated by the pink line. The cone aligns precisely with the center of the CME bubble, demonstrating a good geometric correspondence between the dimming-derived reconstruction and the observed eruption structure. Notably, the upper and lower boundaries of the CME bubble coincide spatially with the secondary dimming regions that develop following the impulsive phase (see the red shaded region in panel (d) of Figure \ref{fig:dimming_nov}). When accounting for these secondary dimmings, the fitted cone fully encompasses the CME structure, as illustrated in Figure \ref{plot:lasco_261111_extended}. Figure \ref{fig:cones_lasco_all} in Appendix \ref{appendix3} shows the projection of best-fit DIRECD cones to LASCO/C2 CME bubbles (at the end of the impulsive phase) for the statistical event sample (excluding events \#1, \#2, \#11, and \#25 due to lack of LASCO/C2 data). Here, events \#4, \#5, \#7, \#8, \#14, \#18, \#19, \#20, \#23, \#24, \#26, and \#32 exhibit low inclination angles ($\beta \leq 10^\circ$), resulting in broader projection spreads. This results from their face-on geometry and enhanced 3D-to-2D projection effects, where the conical structure’s tilt minimizes the apparent angular width. It is evident that the CME direction obtained by DIRECD closely matches the bubble direction observed in the coronagraphs.


\section{Discussion and Conclusion}

This study presents a comprehensive statistical analysis of the DIRECD (Dimming Inferred Estimation of CME Direction) method, which leverages coronal dimmings observed by SDO/AIA to estimate the initial propagation direction of CMEs. By analyzing 33 well-defined on-disk events, we demonstrate that DIRECD provides a robust and observationally constrained approach to determining early CME kinematics before they become fully visible in coronagraphs. This study highlights several key results that demonstrate the effectiveness of the DIRECD method. First, DIRECD-derived inclination angles show strong agreement with GCS model reconstructions, with a mean absolute error $1.2^\circ$ and a standard deviation of differences $\sigma_d = 10.4^\circ$, validating DIRECD as a reliable tool for estimating CME direction in the low corona where traditional coronagraph observations are limited. DIRECD shows that the dominant dimming direction corresponds closely to the CME's principal axis, enabling 3D reconstruction of CME geometry using only EUV dimming observations and providing critical insights into early non-radial motion before significant deflection occurs. Second, projections of best-fit DIRECD cones onto LASCO/C2 images confirm the model accurately captures CME spatial structure near the impulsive phase's end, reinforcing the physical connection between dimmings and CME expansion. Finally, while DIRECD's meridional (latitudinal) angles closely match GCS results (mean absolute error of $0.3^\circ$ and a standard deviation of $7.8^\circ$), longitudinal deviations reveal an eastward bias in GCS reconstructions (mean absolute error of of $-2.9^\circ$, standard deviation of $18.9^\circ$), likely from Parker spiral influences on later-stage propagation \citep{martinic2022determination, martinic2023effects, fargette2022preferential}; DIRECD instead focuses on the initial eruption direction at low coronal heights (up to 3$R_{sun}$) before significant deflection occurs, providing a clearer view of early CME direction unaffected by interplanetary dynamics. We further note that the standard deviation of differences in DIRECD and GCS for longitude/latitude lie within the uncertainties of the GCS method \citep{verbeke2023quantifying}.

The DIRECD method bridges a critical observational gap by providing early CME direction estimates at the end of the impulsive phase of the dimming. This capability is especially valuable for space weather forecasting, where accurate predictions of Earth-directed CMEs depend on understanding their initial trajectory. This method can be effectively combined with established techniques for estimating CME direction at higher altitudes, enabling comprehensive tracking of CME propagation from the low corona to interplanetary space. For instance, it can complement observations from STEREO/COR1, which provide reliable CME direction measurements at heights of $3.5~R_{sun}$ and beyond \citep{mierla2008quick}, as well as the Graduated Cylindrical Shell model, widely used for reconstructing CME morphology and kinematics up to $\sim 20~R_{sun}$ \citep{thernisien2006modeling, thernisien2011implementation}. Another advantage is that the DIRECD method does not need multi-viewpoint observations or coronagraphs. Moreover, DIRECD is compatible with any EUV imager capable of detecting sufficient dimming signatures. It has already been successfully tested using SDO/AIA data at 193 \AA~and 211 \AA~wavelengths, as well as STEREO EUVI 195 \AA~observations. Future applications could integrate DIRECD with near real-time dimming detection algorithms to improve early-warning systems.

\begin{figure} 
	\centering
	\includegraphics[width=0.55\textwidth]{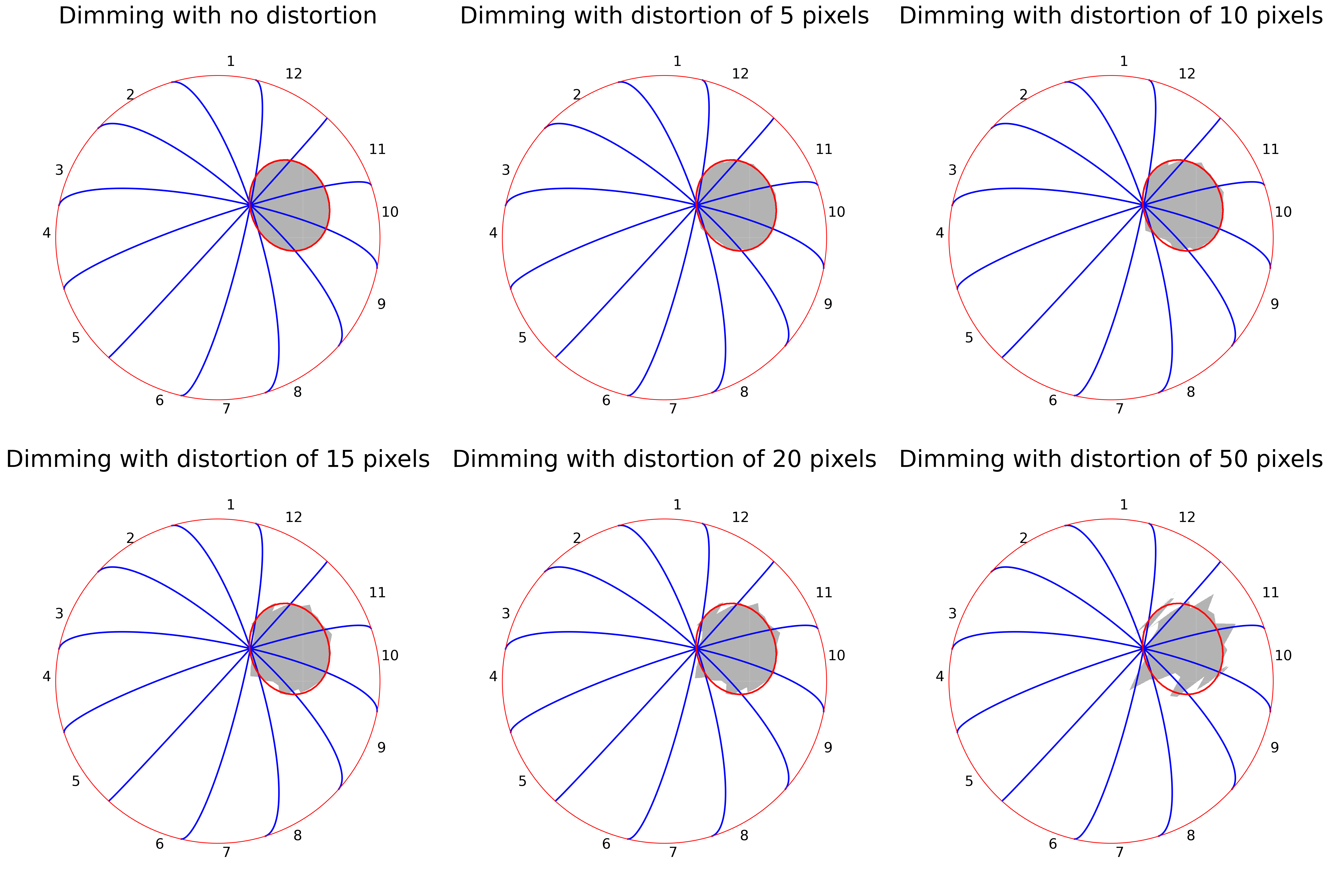}
	\caption{Evolution of dimming profiles with varying noise levels (grey curves), ranging from a standard deviation of 0 (ideal case) to 50 (extremely noisy) pixels. Red curve shows the contour of dimming with 0 noise.} 
	\label{fig:direcd_dimming_limitation}
\end{figure}

\begin{figure}  
	\centering
	\includegraphics[width=0.5\textwidth]{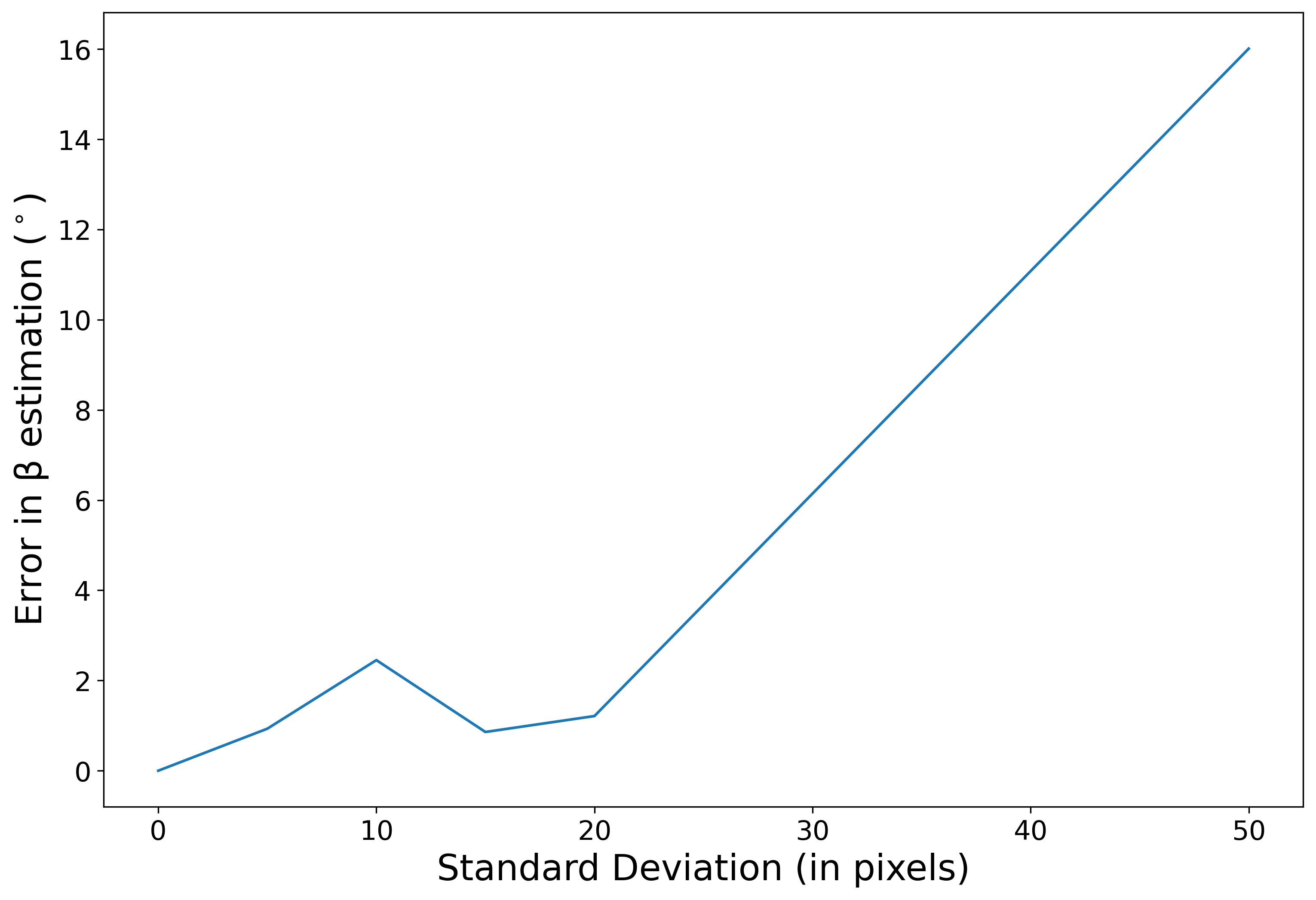}
	\caption{Variation of estimation of inclination angle $\beta$ with standard deviation of pixels.} 
	\label{fig:direcd_beta_statistics}
\end{figure}

However, this method is not without limitations. Firstly, it is currently restricted to well-developed on-disk dimmings observed near the end of the impulsive phase of the dimming. To assess the sensitivity of dimming detection on DIRECD, we analyzed simulated cone projections from \citep{jain2024coronal}, reconstructing the cone geometry from known projections. Using a cone with a height of 1 $R_{sun}$ and an inclination angle of 30$^\circ$, we introduced artificial gaussian noise (0–50 pixels corresponding to 0-60 arcsecs) into the dimming profiles and evaluated its impact on inclination angle ($\beta$) estimation. As shown in Figure \ref{fig:direcd_dimming_limitation}, increasing noise levels introduce distortions in the dimming profiles, while Figure \ref{fig:direcd_beta_statistics} demonstrates that the error in $\beta$ estimation generally tends to increase with noise. However, this relationship is not strictly monotonic - local fluctuations occur due to the complex interplay of factors including choice of source location and random geometric distortions of the dimming structure, which can occasionally lead to transient reductions in estimation error. At a noise level of 50 pixels (corresponding to ~4$^\circ$ latitudinal/longitudinal deviation), the error in $\beta$ reaches 15$^\circ$ (50\% of the true value), underscoring the method’s dependence on precise dimming detection and correct flare-source association. Any inaccuracies in these initial steps may propagate into CME direction estimates. Second, the method is ineffective for stealth CMEs, which lack clear low-coronal signatures \citep{robbrecht2009no,howard2013stealth}. Third, the technique is only valid for the propagation of CMEs within approximately 3 $R_{sun}$, as established by \cite{Dissauer2019}, who found that most dimming evolution occurs within this range. This contrasts with GCS reconstructions, which can track CMEs out to 15-20 $R_{sun}$ and therefore capture later-stage interactions with the solar wind and interplanetary magnetic fields. 

In summary, this study establishes DIRECD as a powerful tool for estimating the early direction of CMEs by linking low-coronal dimmings to the three-dimensional structure of eruptions, offering a complementary approach to existing reconstruction techniques such as GCS at higher coronal altitudes. The consistent validation against both coronagraph observations and independent 3D models underscores DIRECD’s potential to advance space weather forecasting. Work is ongoing to develop DIRECD-soft, a user-friendly GUI, and the code will be made publicly available on GitHub to facilitate broader adoption and further validation by the community.

\begin{acknowledgments}
SDO data is courtesy of NASA/SDO and the AIA, and HMI science teams. The Large Angle Spectroscopic Coronagraph (LASCO) is a three coronagraph package which has been jointly developed for the Solar and Heliospheric Observatory (SOHO) mission by the Naval Research Laboratory (USA), the Laboratoire d'Astronomie Spatiale (France), the Max-Planck-Institut für Aeronomie (Germany), and the University of Birmingham (UK). We thank the referee for their valuable comments.
\end{acknowledgments}

\begin{contribution}

S.J. and T.P. developed the method, with S.J. leading the writing of the paper. A.V., K.D., and A.R. contributed to the conceptualization of this study, data analysis, and writing. All authors discussed the results and provided feedback on the manuscript.




\end{contribution}

%

\software{Python/Sunpy \citep{mumford2015sunpy}, IDL/Solarsoft \citep{freeland1998data}}

\clearpage
\appendix

\section{Cone Construction}\label{appendix1}

\begin{figure}[h]
\centering 
\subfloat[]{%
  \includegraphics[width=0.35\columnwidth]{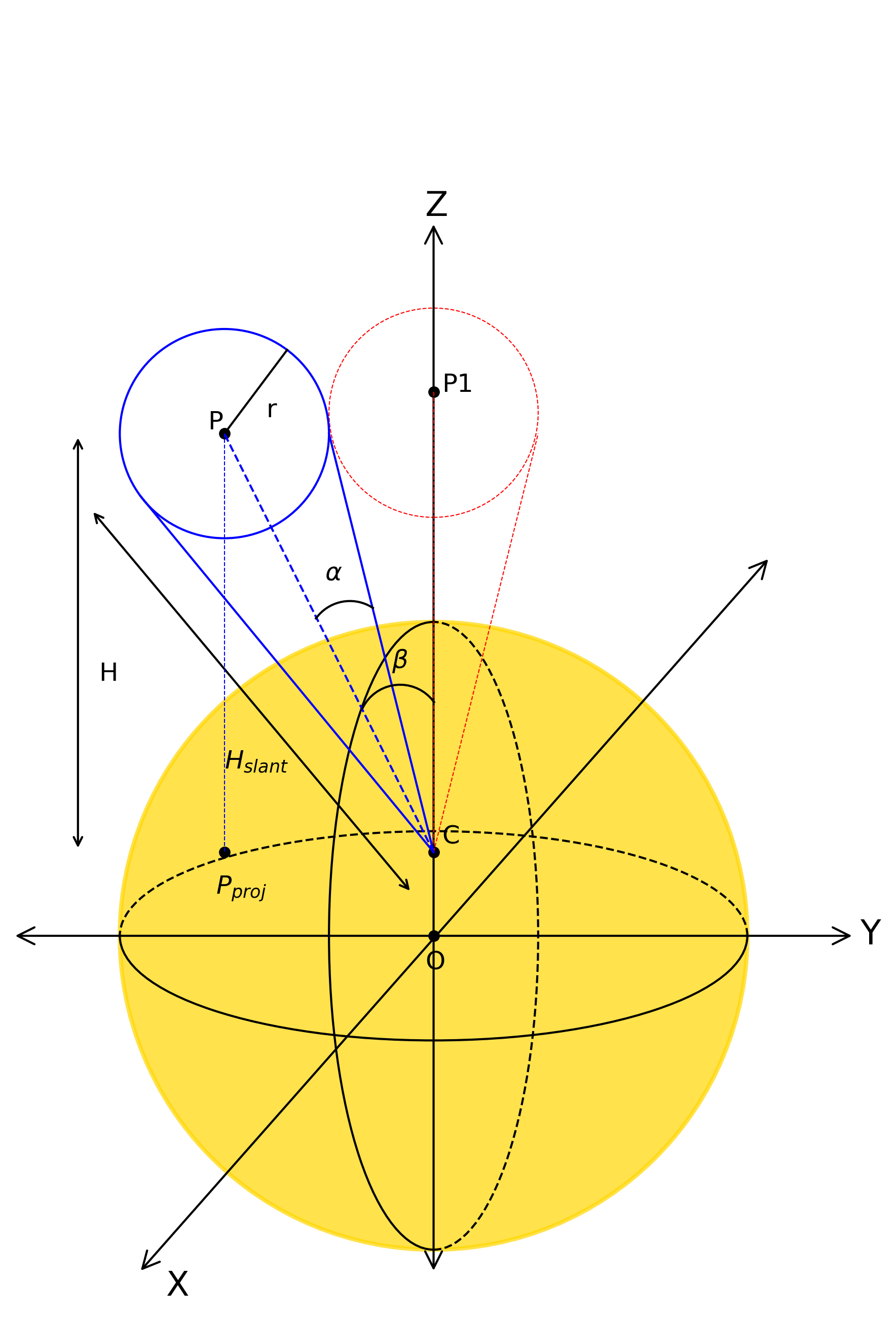}%
  \label{plot:2d_model}%
}\qquad
\subfloat[]{%
  \includegraphics[width=0.5\columnwidth]{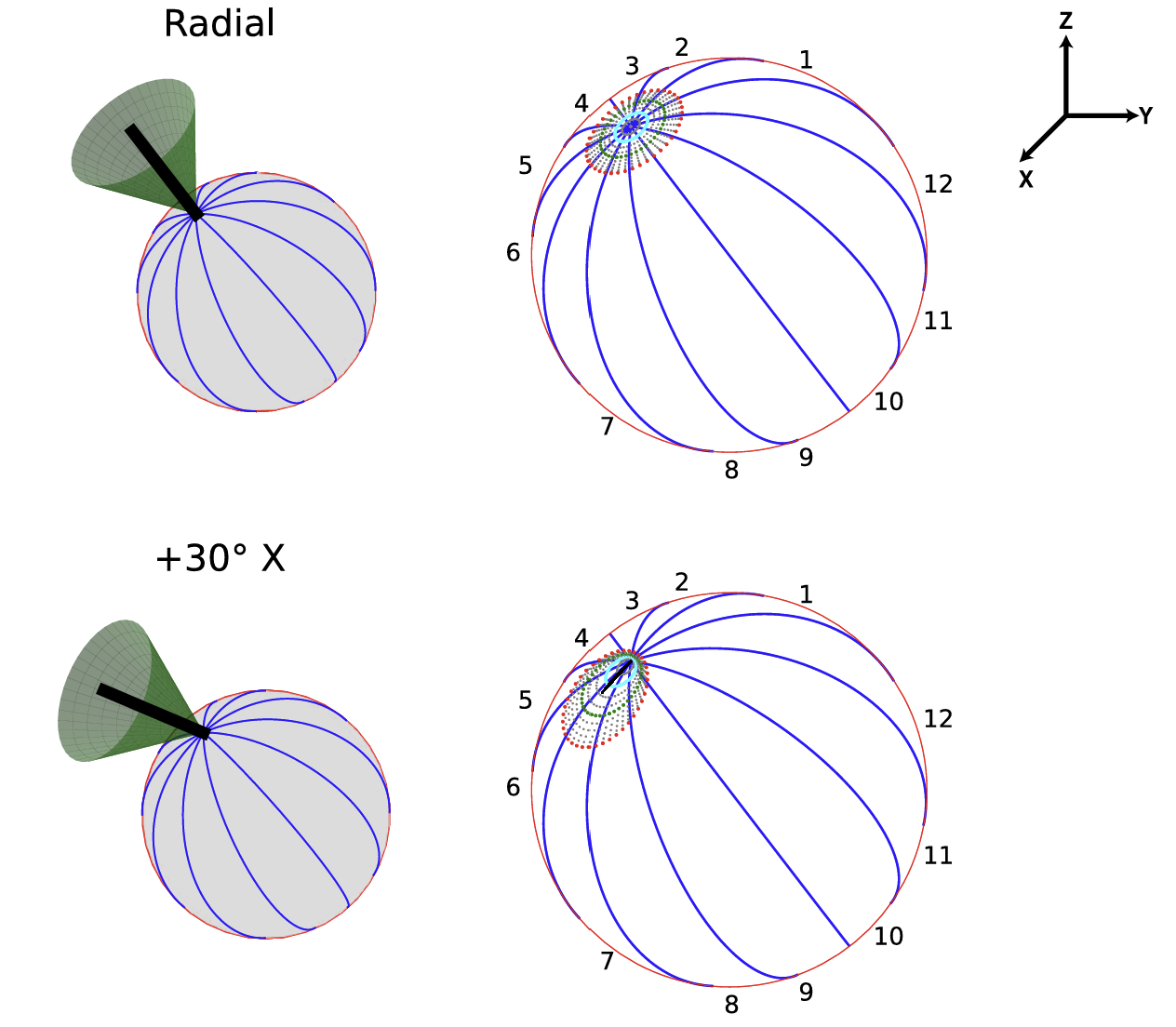}%
  \label{plot:proj_example}%
}
\caption{Left Panel: Schematic of the cone geometry in 2D. The red cone is aligned radially, while the blue cone is tilted by an angle $\beta$ relative to the radial axis. The half-angle of the cone is $\alpha$, with C denoting the source position and O marking the solar center. H is the perpendicular height of point P above its surface projection ($P_{proj}$), and $H_{slant}$ indicates the slant height of the cones. Right Panel: 3D configuration of radial and inclined cones along with their orthogonal projections for a cone height of $1~R_{sun}$. The cyan, green, and red projections correspond to cone heights of $0.2$, $0.5$, and $1$ $R_{sun}$, respectively.}
\end{figure}

Figure \ref{plot:2d_model} illustrates a 2D schematic of the cone structure. The red cone represents the radially oriented cone, while the blue cone is inclined at an angle $\beta$ relative to the radial direction. The parameter $\alpha$ denotes the half-width of the cone, point C marks the source location, and O indicates the Sun’s center. The perpendicular height of point P from its orthogonal projection ($P_{proj}$) on the surface is denoted as H, and $H_{slant}$ represents the slant height of the cones.

In Figure \ref{plot:proj_example}, we display the 3D configuration of both the radial and inclined cones, along with their orthogonal projections (shaded in grey) onto the solar surface for a cone height of $1~R_{sun}$. The cyan, green, and red projections correspond to cone heights of $0.2$, $0.5$, and $1$ $R_{sun}$, respectively.

The orthogonal projections of the cone on the solar surface can be described by the following equation:
\begin{align}
\label{ortho}
X_{ortho} = \frac{R_{sun} \cdot X_c}{\sqrt{X_c^2 + Y_c^2 + Z_c^2}} \nonumber \\
Y_{ortho} = \frac{R_{sun} \cdot Y_c}{\sqrt{X_c^2 + Y_c^2 + Z_c^2}} \\
Z_{ortho} = \frac{R_{sun} \cdot Z_c}{\sqrt{X_c^2 + Y_c^2 + Z_c^2}} \nonumber.
\end{align}

Here, $X_{ortho}$, $Y_{ortho}$, and $Z_{ortho}$ denote the Cartesian coordinates of the orthogonal projections, while $X_c$, $Y_c$, and $Z_c$ represent the three-dimensional coordinates of the cone. The parameter $R_{sun}$ corresponds to the solar radius in kilometers. The height H is defined as:

\begin{equation}\label{cone_height}
H = \sqrt{(P_{x} - P^{proj}_{x})^2 + (P_{y} - P^{proj}_{y})^2 + (P_{z} - P^{proj}_{z})^2}
\end{equation}
where ($P_{x}, P_{y}, P_{z}$) and ($P^{proj}_{x},~P^{proj}_{y},~ P^{proj}_{z}$) are the (x, y, z) coordinates of the cone center in space $P$ and its orthogonal projections on the surface.

The slant height of the cone $H_{slant}$ is: 

\begin{equation}\label{cone_slant}
H_{slant} = \sqrt{(P_{x} - C_{x})^2 + (P_{y} - C_{y})^2 + (P_{z} - C_{z})^2}
\end{equation}

where ($C_{x}, C_{y}, C_{z}$) and ($P_{x}, P_{y}, P_{z}$) are the (x, y, z) coordinates of the flare source $C$. 

The radius of the base of the cone r is related to slant height $H_{slant}$ as:
\begin{gather}
H_{slant} = r\cdot tan~\alpha
\end{gather}

The inclination angle $\beta$ is given by:
\begin{equation}\label{inclination_angle}
\beta = \arccos\left( \frac{\vec{v}_1 \cdot \vec{v}_2}{\|\vec{v}_1\| \|\vec{v}_2\|} \right)
\end{equation}
where $\vec{v}_1$ is the vector between Sun center and $C$ and $\vec{v}_2$ is the vector between $C$ and $P$.

\clearpage

\section{Cone projections}\label{appendix2}

\begin{figure}[h]
	\centering
	\includegraphics[width=0.85\textwidth]{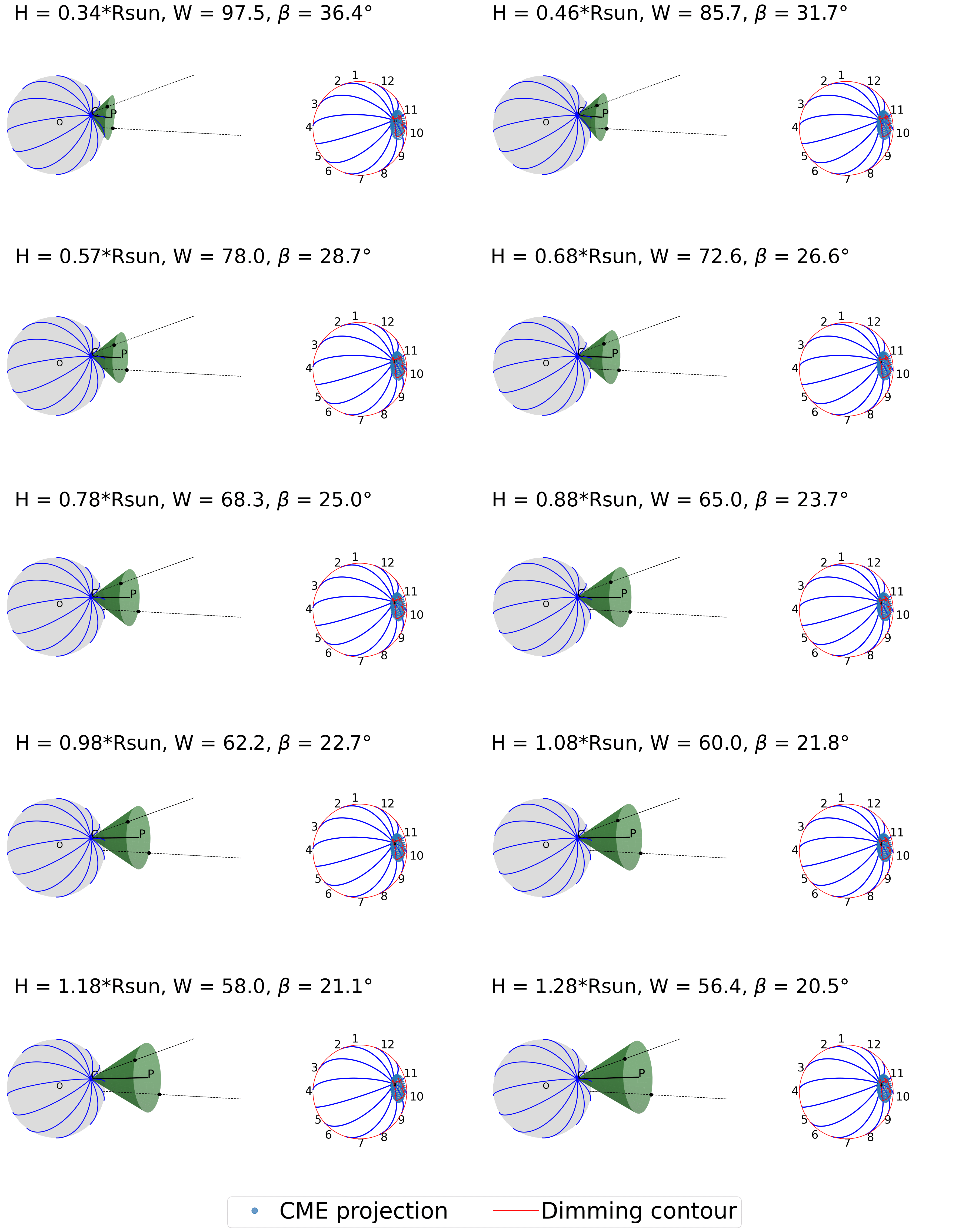}
	\caption{The panels are arranged in lexicographical order (left-to-right then top-to-bottom), showing} CME cones at heights of 0.34–1.28~$\rm{R_{sun}}$ with associated widths of 97.5$^\circ$-56.4$^\circ$ and inclination angles of 36.4$^\circ$ - 20.5$^\circ$ (Cols. 1 and 3) and their orthogonal projections (blue dots) onto the solar sphere (Cols. 2 and 4) for the November 26, 2011 event. The dimming boundaries are outlined in red.
	\label{fig:cone_projection_1}
\end{figure}

\begin{figure}[h]
	\centering
	\includegraphics[width=0.9\textwidth]{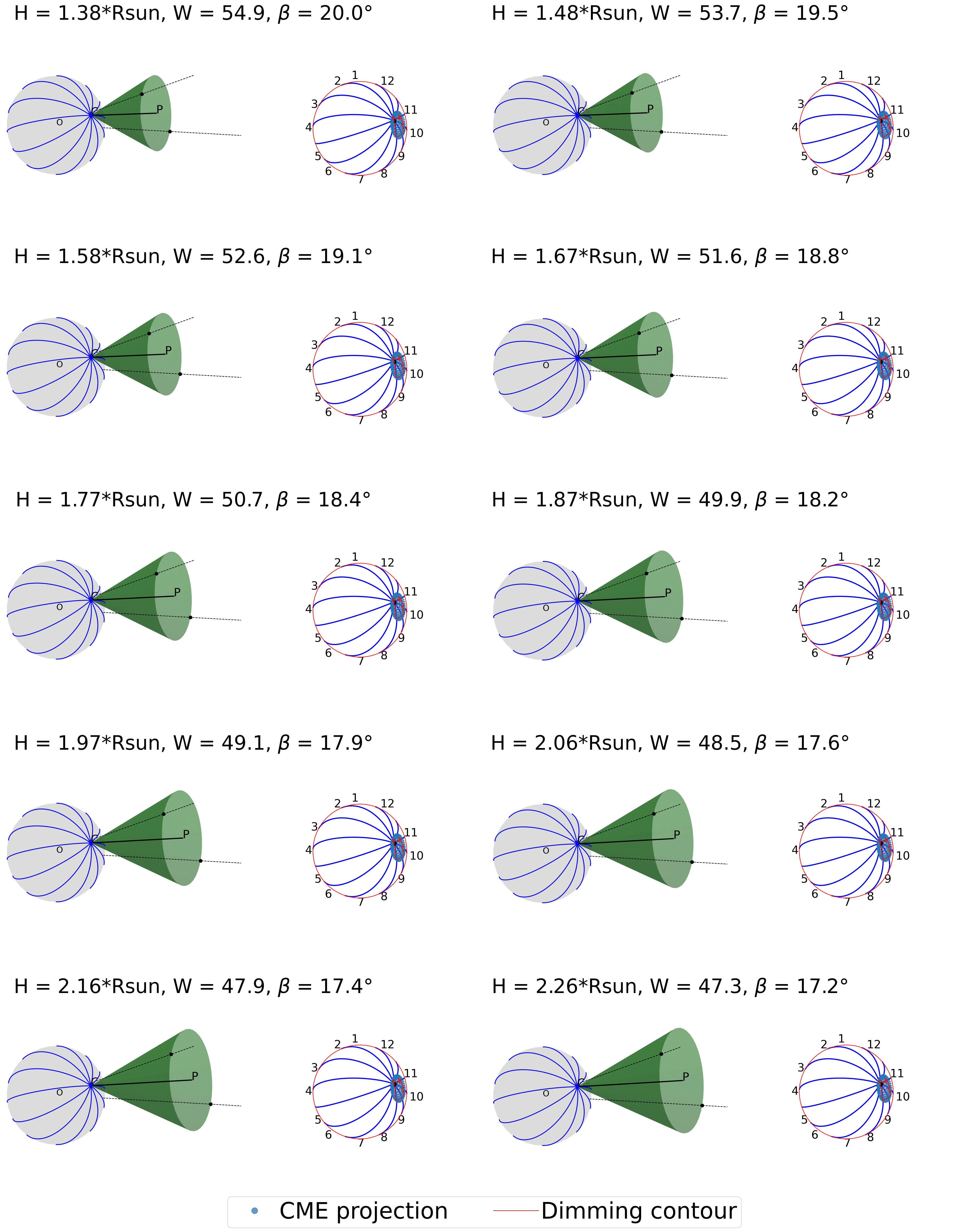}
	\caption{Same as Figure \ref{fig:cone_projection_1} for CME cones at heights of 1.38–2.26~$\rm{R_{sun}}$ with associated widths of 54.9$^\circ$-47.3$^\circ$ and inclination angles of 20$^\circ$ - 17.2$^\circ$.} 
	\label{fig:cone_projection_2}
\end{figure}
\clearpage

\section{LASCO/C2 VALIDATION}\label{appendix3}

\begin{figure}[h]
	\centering
	\includegraphics[width=0.85\textwidth]{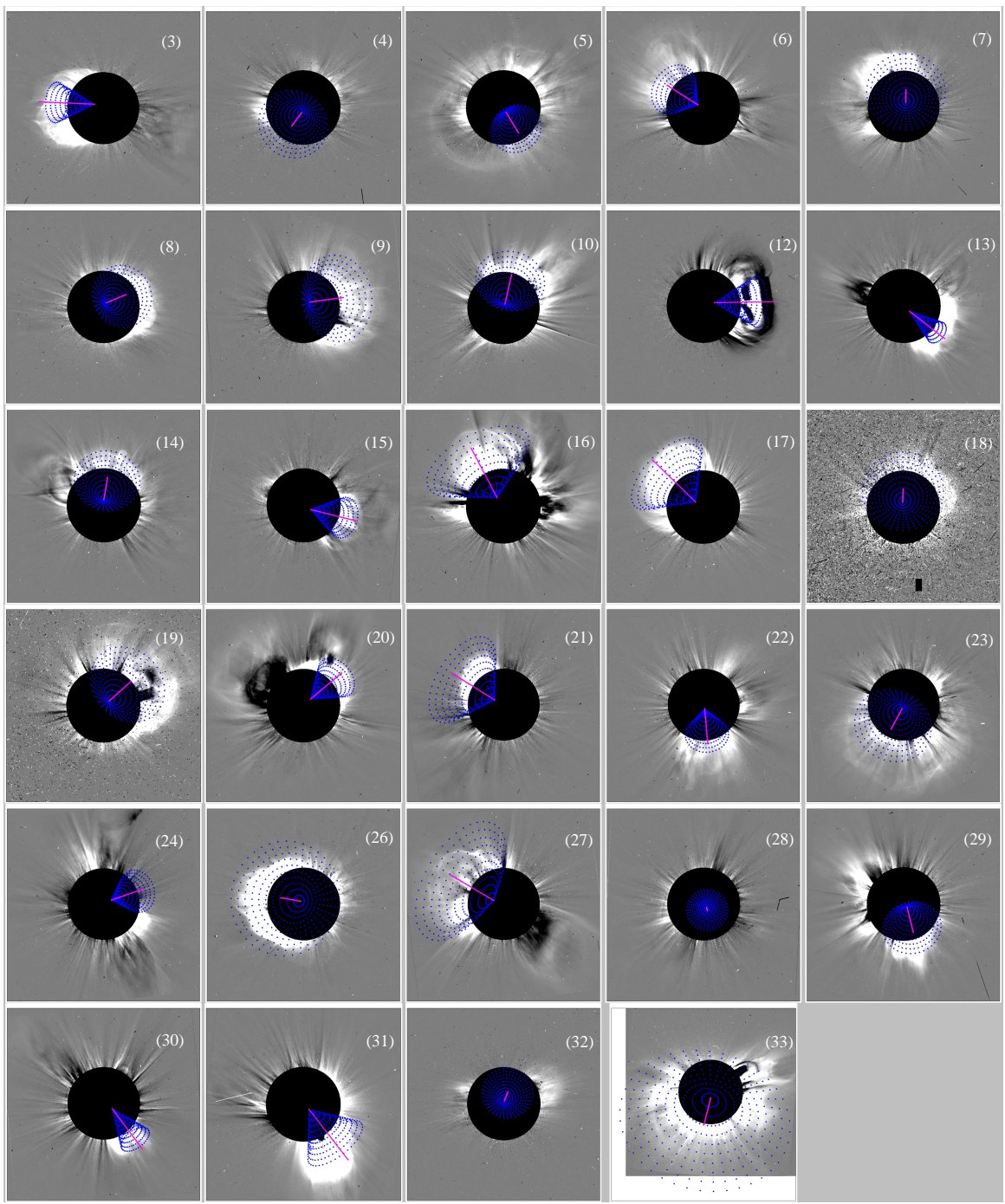}
	\caption{Composite view of CME bubbles observed by the LASCO/C2 coronagraph (at the end of the impulsive phase) for the statistical event sample (excluding events \#1, \#2, \#11, and \#25 due to lack of coronagraph data). Blue dots represent the best-fit DIRECD cone model projections, while the pink line indicates the central CME propagation axis. Events \#4, \#5, \#7, \#8, \#14, \#18, \#19, \#20, \#23, \#24, \#26, and \#32 exhibit inclination angles ($\beta \leq 10^\circ$), resulting in broader projection spreads. This results from their face-on geometry and enhanced 3D-to-2D projection effects, where the conical structure’s tilt minimizes the apparent angular width.} 
	\label{fig:cones_lasco_all}
\end{figure}


\clearpage
\bibliography{ms}{}
\bibliographystyle{aasjournal}



\end{document}